\documentclass[aps,prd,preprint,superscriptaddress,preprintnumbers]{revtex4-1}

\pdfoutput=1
\usepackage{mathrsfs}
\usepackage{amsfonts}
\usepackage{amsmath,amssymb,bm,bbm}
\usepackage{physics} 
\usepackage[colorlinks=true,linkcolor=blue,citecolor=blue,urlcolor=blue]{hyperref}
\usepackage{epsfig}
\usepackage{graphicx}
\usepackage{slashed}
\usepackage{booktabs}
\usepackage{tabu}
\usepackage[dvipsnames]{xcolor}
\usepackage[normalem]{ulem}
\usepackage{tikz}
\usepackage{soul}
\usetikzlibrary{shapes,arrows}
\usetikzlibrary{positioning}
\usepackage{ulem}

\newcommand{\muL}{{\mu_{\rm lat}}}
\newcommand{\muLsq}{{\mu_{\rm lat}^2}}

\begin{document}
\preprint{JLAB-THY-22-3592}

\title{Complementarity of experimental and lattice QCD data \\ on pion parton distributions}

\author{P. C. Barry}
\affiliation{Jefferson Lab,
        Newport News, Virginia 23606, USA}
\author{C. Egerer}
\affiliation{Jefferson Lab,
        Newport News, Virginia 23606, USA}
\author{J. Karpie}
\affiliation{\mbox{Physics Department, Columbia University,
        New York City, New York 10027, USA}}
\author{W. Melnitchouk}
\affiliation{Jefferson Lab,
        Newport News, Virginia 23606, USA}
\author{C. Monahan}
\affiliation{Jefferson Lab,
        Newport News, Virginia 23606, USA}
\affiliation{\mbox{Department of Physics, William \& Mary,
        Williamsburg, Virginia 23185, USA}}
\author{K. Orginos}
\affiliation{Jefferson Lab,
        Newport News, Virginia 23606, USA}
\affiliation{\mbox{Department of Physics, William \& Mary,
        Williamsburg, Virginia 23185, USA}}
\author{Jian-Wei Qiu}
\affiliation{Jefferson Lab,
        Newport News, Virginia 23606, USA}
\affiliation{\mbox{Department of Physics, William \& Mary,
        Williamsburg, Virginia 23185, USA}}
\author{D. Richards}
\affiliation{Jefferson Lab,
        Newport News, Virginia 23606, USA}
\author{N. Sato}
\affiliation{Jefferson Lab,
        Newport News, Virginia 23606, USA}
\author{R. S. Sufian}
\affiliation{Jefferson Lab,
        Newport News, Virginia 23606, USA}
\affiliation{\mbox{Department of Physics, William \& Mary,
        Williamsburg, Virginia 23185, USA}}
\author{S. Zafeiropoulos}
\affiliation{Aix Marseille Univ, Universit\'{e} de Toulon,
        CNRS, CPT, Marseille, France\\
        \vspace*{0.2cm}
        {\bf Jefferson Lab Angular Momentum (JAM) and HadStruc Collaborations
        \vspace*{0.2cm} }}

\begin{abstract}
We extract pion parton distribution functions (PDFs) in a Monte Carlo global QCD analysis of experimental data together with reduced Ioffe time pseudo-distributions and matrix elements of current-current correlators generated from lattice QCD.
By including both experimental and lattice QCD data, our analysis rigorously quantifies both the uncertainties of the pion PDFs and systematic effects intrinsic to the lattice QCD observables.
The reduced Ioffe time pseudo-distributions significantly decrease the uncertainties on the PDFs, while the current-current correlators are limited by the systematic effects associated with the lattice.
Consistent with recent phenomenological determinations, the behavior of the valence quark distribution of the pion at large momentum fraction is found to be $\sim (1-x)^{ \beta_{\rm eff}}$ with $\beta_{\rm eff} \approx 1.0-1.2$.
\end{abstract}

\date{\today}
\maketitle

%%%%%%%%%%%%%%%%%%%%%%%%%%%%%%%%%%%%%%%%%%%%%%%%%%%%%%%%%%%%%%%%%%%%%%
\section{Introduction}

The pion is one of the most important particles in nature, yet three-quarters of a century after its discovery fundamental questions remain about its properties and behavior.
On the one hand, it has long been identified as the pseudo-Goldstone boson associated with chiral symmetry breaking, which governs the low-energy interactions of hadrons.
On the other hand, the pion's internal structure in terms of its bound state of quarks and gluons (partons) can be revealed in high-energy scattering reactions.
Understanding the structure of the pion provides insight into the nature of the strong force itself.

Details of the internal partonic structure of the pion have remained elusive because it cannot be probed as a fixed target.
However, secondary pion beams have been used to scatter from stationary nuclear targets, and inclusive Drell-Yan (DY) lepton-pair production measurements have been made at CERN and Fermilab~\cite{Conway:1989fs, NA10:1985ibr}.
These, along with prompt photon data, have been used in phenomenological QCD analyses to extract the pion parton distribution functions (PDFs) at moderate and large parton momentum \mbox{fractions, $x$ \cite{Owens:1984zj, Aurenche:1989sx, Sutton:1991ay, Gluck:1991ey, Gluck:1999xe, Wijesooriya:2005ir, Novikov:2020snp}}.
Recently, the Jefferson Lab Angular Momentum (JAM) collaboration utilized the leading neutron (LN) deep-inelastic electroproduction measurements from HERA \cite{H1:2010hym, ZEUS:2002gig}, together with the DY data, to capture the small momentum fraction region and constrain the sea quark and gluon distributions in the pion~\cite{Barry:2018ort}.
Further studies in this vein have been performed with the addition of large-$p_T$ differential DY data \cite{Cao:2021aci}, with its enhanced sensitivity to the gluon distribution, as well as with threshold resummation on the DY hard coefficients \cite{Barry:2021osv}.

The behavior of the valence quark distribution in the region where the momentum fraction is large has elicited much debate, particularly in the limit where $x\to 1$.
Here, the valence PDF is modeled by an asymptotic behavior $\sim (1-x)^{\beta_{\rm eff}}$, and various nonperturbative
and perturbative QCD models predict different values for $\beta_{\rm eff}$, ranging from $\beta_{\rm eff}\approx 1$ to $\beta_{\rm eff}\approx 2$ \cite{Ezawa:1974wm, Landshoff:1973pw, Gunion:1973ex, Farrar:1979aw, Berger:1979du, Shigetani:1993dx, Szczepaniak:1993uq, Davidson:1994uv, Hecht:2000xa, Melnitchouk:2002gh, Noguera:2015iia, Hutauruk:2016sug, Hobbs:2017xtq, deTeramond:2018ecg, Bednar:2018mtf, Lan:2019vui, Lan:2019rba, Chang:2020kjj, Cui:2020tdf, Kock:2020frx,Cui:2022bxn}.
Experiments have been proposed to further constrain the large-$x$ behavior using the Sullivan process at Jefferson Lab~\cite{TDISProposal} and the future Electron-Ion Collider (EIC)~\cite{AbdulKhalek:2021gbh, Arrington:2021biu}.
An earlier analysis using threshold resummation on the DY data found $\beta_{\rm eff} \approx 2$ \cite{Aicher:2010cb}, while more recently the JAM collaboration performed a global analysis of pion PDFs using more advanced threshold resummation technology, particularly the double Mellin method~\cite{Westmark:2017uig}, finding $\beta_{\rm eff} \approx 1.2$~\cite{Barry:2021osv}.

Experimental cross sections are related to the integrals of PDFs from $x_{\rm min}$ to 1 with $x_{\rm min}$ fixed by external kinematics.
In fact, no experiment can measure the parton momentum fraction $x$ directly.
If the kinematics force $x_{\rm min}$ to be $\sim 1$ or near the exclusive threshold, the role of logarithms in coefficient functions that match the PDFs to cross sections becomes important.
Consequently, PDFs extracted from these cross sections become highly sensitive to the perturbative order of the matching coefficients, particularly the threshold effects.
In Ref.~\cite{Barry:2021osv}, threshold resummation stemming from soft gluon radiation was applied to the DY hard coefficients up to next-to-leading logarithmic (NLL) accuracy as a potentially necessary theoretical correction to the perturbative expansion.
Since the PDFs fall rapidly near threshold, the resummed hard coefficients may contribute substantially to the overall hadronic cross section \cite{Aicher:2010cb, Westmark:2017uig, Shimizu:2005fp, Catani:1996yz}.
Threshold resummation was not applied in \cite{Barry:2021osv} to the LN cross sections since DIS-like observables have less of an overall impact from large logarithms and are less sensitive to large-$x$ behaviors than for the case of  DY~\cite{Sterman:2000pu}.
The description of both the DY and LN data was compatible with the inclusion of the double Mellin resummation \cite{Barry:2021osv}, so both the next-to-leading order (NLO) and NLO+NLL methods of calculating the short distance coefficients remain viable alternatives for extracting PDFs from available experimental data.

On the theoretical side, there has also been considerable recent interest in computing the pion's internal structure through lattice QCD simulations~\cite{Zhang:2018nsy,Fan:2018dxu,Sufian:2019bol,Chen:2019lcm, Izubuchi:2019lyk, Joo:2019bzr, Lin:2020ssv,Sufian:2020vzb, Karthik:2021qwz,Fan:2021bcr}.
Unlike the experimental cross sections, lattice QCD--calculable matrix elements of operators composed of two-quark, two-gluon or two-current correlations with a separation $z$ in position space are matched to PDFs in terms of the short-distance factorization (SDF) formalism through integrals over the entire range of $x$~\cite{Ma:2014jla,Radyushkin:2017cyf, Ma:2017pxb}.
The extraction of PDFs from these matrix elements is therefore not as sensitive to threshold resummation of the corresponding matching coefficients as it is in the case of DY~\cite{Gao:2021hxl}. 
The sensitivity to the large-$x$ behavior of the PDFs depends on the relative size of the large-$x$ contribution to the matching coefficient at different values of $z$.
Like the experimental cross sections, these matrix elements are also subject to power corrections from the large-$z$ regime, which limits the lattice data that can be used for extracting PDFs in the SDF approach. 
On the other hand, in the large momentum effective theory (LaMET) approach, lattice data for the same two-parton correlation matrix elements are taken at all $z$ values and are Fourier transformed into the quasi-PDF in momentum space, which is then matched to the PDF as the pion momentum $P_z \to \infty$.
For finite $P_z$ values, the matching between the quasi-PDF and PDF also has power corrections that are proportional to $[x^2 (1-x) P_z^2]^{-1}$, which could hamper the ability to accurately determine the large-$x$ behavior of the PDFs~\cite{Braun:2018brg}.

The LaMET and SDF approaches provide complementary methods for extracting PDFs from lattice calculations, with different systematic uncertainties. 
The success of these approaches has encouraged significant effort aimed at constraining PDFs through lattice QCD, using a variety of methods and lattice actions~\cite{Braun:2007wv, Chambers:2017dov, Alexandrou:2018pbm, Lin:2018pvv, Bhat:2020ktg, Alexandrou:2018eet, Joo:2019jct, Izubuchi:2019lyk, Fan:2020nzz, Joo:2020spy, Cichy:2019ebf, DelDebbio:2020rgv, Detmold:2021uru, Egerer:2021ymv, HadStruc:2021wmh, Egerer:2021dgg}.
These methods generally provide more information about PDFs than does the traditional approach of calculating the Mellin moments of PDFs from matrix elements of local operators. 
That approach has been limited to the lowest few moments by the reduced rotational symmetry of the Euclidean lattice, and generally provides weaker constraints on PDFs than methods that give access to the $x$ dependence directly~\cite{Detmold:2003tm, Detmold:2003rq, Lin:2017snn}.

In the present analysis, we make use of new data from lattice QCD simulations of pion matrix elements with space-separated quark-quark and current-current (CC) correlators, which have well-defined continuum limits, and can be factorized to the standard light-cone PDFs with perturbatively calculable matching coefficients. 
The lattice data on matrix elements of quark-quark correlators are presented in the form of reduced Ioffe time pseudo-distributions (Rp-ITDs)~\cite{Radyushkin:2017cyf, Karpie:2018zaz}. 
In this analysis lattice data on both Rp-ITDs and matrix elements of CC correlators are treated on the same footing as experimental data, and are analyzed simultaneously.
In the JAM framework, both experimental and lattice data have been used to constrain the unpolarized and polarized nucleon PDFs~\cite{Bringewatt:2020ixn}, as well as the nucleon's transversity distribution~\cite{Lin:2017stx}. 
The complementarity of the experimental and lattice observables provides an opportunity to learn about both PDFs and the systematic effects associated with lattice calculations of hadron structure.

We proceed with the organization of this paper as follows.
In Sec.~\ref{s.framework} we outline the analysis framework, describing both the experimental and lattice observables, and summarizing the QCD analysis methodology employed here.
The results of the simultaneous analysis of experimental and lattice QCD data are presented in Sec.~\ref{s.results}.
There, we discuss in detail the impact of the lattice Rp-ITDs and CC correlator data on the pion valence quark PDF, and in particular the quantification of the systematic uncertainties, for both the NLO and NLO+NLL approximations to the hard coefficients.
Finally, in Sec.~\ref{s.conclusions} we summarize our conclusions and discuss  future directions in which this analysis can be taken.

%%%%%%%%%%%%%%%%%%%%%%%%%%%%%%%%%%%%%%%%%%%%%%%%%%%%%%%%%%%%%%%%%%%%%%
\section{Analysis framework} \label{s.framework}

Confinement ensures that quarks and gluons cannot be directly observable.
Instead, theoretical and analytic tools are needed to infer the internal partonic structure of hadrons from ``good'' cross sections, defined as those that can be factorized in QCD into convolutions of soft, universal nonperturbative parts, such as those involving PDFs, and perturbatively calculable, short-distance hard coefficients.
Corrections to these factorized theoretical cross sections are suppressed by powers of the large momentum transfer involved in the scattering process. 
Predictions follow when cross sections with different hard coefficients, but the same nonperturbative parts, are compared.

While experiments do not detect partons themselves, the detection of hadrons in the final state can reveal the initial-state structure of the hadronic target from the experimental data through Bayesian inference. 
Such tools allow us to build theoretical observables describing processes involving partons, and to compare them with the available data.
Much in the same way as different experimental datasets are used in phenomenological extractions of PDFs, the universality of the PDFs in different lattice observables can also be tested.
Here, we treat lattice QCD data on the same footing as experimental data.
We use factorization theorems for lattice observables to describe the results in terms of convolutions of hard coefficients with PDFs, and the approach to the extraction of PDFs from lattice data is analogous to the methods used in global QCD analysis of experimental data.
In practice, in our analysis we include DY and LN electroproduction observables in conjunction with Rp-ITD and CC correlator lattice QCD data to extract the PDFs in the pion.
In the following, we describe the theoretical and analysis framework employed in this analysis.

%.....................................................................
\subsection{Experimental observables}

%--Drell-Yan
In the DY process \cite{Drell:1970wh} two hadrons collide with an invariant center of mass energy $\sqrt{S}$, producing a detected $\mu^+ \mu^-$ lepton pair with invariant mass $Q$.
The cross section can be factorized as a convolution of the PDFs of the incident pion and the target nucleus $A$, from which the pion scatters, with the perturbatively calculable hard coefficients.
Reported differentially in the Feynman variable,
    $x_F = x_\pi^0 - x_A^0$,
and in 
    $\sqrt{\tau} = Q/\sqrt{S}$,
the cross section is written as \cite{Collins:1983ju}
\begin{equation}
\frac{\dd{\sigma}}{\dd{x_F}\dd{\sqrt{\tau}}}
= \frac{4\pi \alpha^2}{9\, Q^2 S} 
\sum_{ij} 
\int_{x_\pi^0}^1 \dd{x_\pi}
\int_{x_A^0}^1   \dd{x_A}\,
f_i^\pi(x_\pi,\mu)\, f_j^A(x_A,\mu)\, 
\mathcal{C}^{\rm \tiny{DY}}_{ij}(x_\pi,x_\pi^0,x_A,x_A^0,Q,\mu),
    \label{eq.DY}
\end{equation}
where $f_{i(j)}^{\pi (A)}$ is a PDF of flavor $i~(j)$ in $\pi~(A)$.
The variable 
    $x_{\pi(A)}^0 = \sqrt{\tau}\, e^{\pm Y}$
is the minimum momentum fraction to produce the lepton pair, where $Y$ is the rapidity of the lepton pair.

The hard coefficient $\mathcal{C}^{\rm \tiny{DY}}$ is calculated through perturbative QCD up to a desired order in the strong coupling $\alpha_s$.
We use two forms of the calculation of the short-distance coefficient in DY: 
 (i) fixed order up to $\mathcal{O}(\alpha_s)$, referred to as ``NLO'', and 
(ii) fixed order with threshold resummation up to NLL accuracy, which we refer to as ``NLO+NLL''.
The threshold region corresponds to values of 
    $z = \tau / x_\pi x_A \to 1$,
and large logarithms appear in the form
    $\alpha_s^k \log^{2k-1}(1-z)/(1-z)$
for all orders $k$.
In some regions of $x_F$ and $\tau$, the soft gluon resummation greatly influences $\mathcal{C}^{\rm \tiny{DY}}$, leading to significant differences between the hard coefficients calculated at NLO and NLO+NLL.
These changes are reflected in the resulting PDFs, and the universality of these PDFs must be tested by independent observables.
In Ref.~\cite{Barry:2021osv}, several methods of calculating the threshold resummation were studied, based on the ``Mellin-Fourier'' and ``double Mellin'' methods.
The Mellin-Fourier approach was shown by Lustermans {\it et al.}~\cite{Lustermans:2019cau} to miss certain terms in the resummation of the same logarithmic orders, whereas the double Mellin method accounted for these corrections~\cite{Westmark:2017uig}.
The Mellin-Fourier approaches also showed a worse description of the data at large $x_F$ and $\tau$.
In this analysis, we use the preferred double Mellin approach~\cite{Westmark:2017uig} to compute the NLO+NLL result and compare it with that obtained from the pure NLO calculation.

The pion-induced DY data were taken from the Fermilab E615 experiment~\cite{Conway:1989fs} and the NA10 experiment~\cite{NA10:1985ibr}  at CERN.
The E615 experiment made use of a pion beam with an energy of 252~GeV, while the NA10 experiment utilized beam energies of 194~GeV and 286~GeV.
To avoid the $J/\psi$ and $\Upsilon$ resonance regions and edges of phase space we limit ourselves to the range $4.16 < Q < 7.68$~GeV and $0 < x_F < 0.9$.
After these kinematic cuts are made, we fit to 61 data points from the E615 experiment, and 36 and 20 points from the NA10 experiment with the 194~GeV and 286~GeV pion beam energies, respectively.
Since the data were collected on a tungsten target, tungsten PDFs need to be used in the calculation, and for these we take the central values from the EPPS16~\cite{Eskola:2016oht} global nuclear PDF analysis.
We find, however, that our results are insensitive to the changes of the tungsten PDFs between the EPPS16~\cite{Eskola:2016oht} and nCTEQ~\cite{Kovarik:2015cma} nuclear PDF parametrizations.

%--Leading neutron
The second method of obtaining information on the structure of the pion is less direct but also relatively well established.
In this case, LN electroproduction is used through the Sullivan process \cite{Sullivan:1971kd} to probe the pion structure function in tagged deep-inelastic scattering, $e p \to e' n X$, with the neutron detected in the far forward region.
In the one-photon exchange approximation, the electron beam radiates a virtual photon with 4-momentum $q=\ell -\ell'$, where $\ell~(\ell')$ is the initial (final) state lepton momentum, and $q^2 \equiv -Q^2$, while the momentum transfer squared between the initial and final nucleons is $t=(p-p')^2$, with $p~(p')$ the 4-momentum of the initial (final) state nucleon.
The LN cross section is then proportional to the LN structure function,
\begin{equation}
    \frac{\dd^3{\sigma}}{\dd{x}\dd{Q^2}\dd{x_L}} = \frac{4\pi \alpha^2}{xQ^4} \left(1-y_e + \frac{y_e^2}{2} \right) F_2^{\rm LN}(x,Q^2,x_L),
\end{equation}
where $x=Q^2/2p\cdot q$ is the Bjorken scaling variable, $x_L$ is the longitudinal momentum fraction the neutron carries relative to the initial proton, $y_e=q\cdot p/q\cdot \ell \approx Q^2/sx$ is the lepton inelasticity, and $s=(\ell+p)^2$ is the total invariant mass squared of the collision.
In the limit $|t| \to 0$, which occurs in the region $x_L \to 1$, the virtual photon absorption reaction $\gamma^* p \to n X$ is dominated by the exchange of pions~\cite{Sullivan:1971kd, Thomas:1983fh, Melnitchouk:1992yd, DAlesio:1998uav, Kopeliovich:2012fd}, which can be described through the $p \to \pi^+ n$ splitting function, $f_{\pi N}$.
The LN structure function is represented as a convolution of $f_{\pi N}$ with the structure function of the pion, $F_2^\pi$,
\begin{equation}
F_2^{{\rm LN}}(x,Q^2,x_L) = 2 f_{\pi N}(\bar{x}_L)\, F_2^\pi(x_\pi,Q^2),
    \label{eq.F2LN}
\end{equation}
where the splitting function $f_{\pi N}$, evaluated at $\bar{x}_L \equiv 1-x_L$, is the light-cone momentum distribution of the pion in the nucleon whose functional form can be found in Refs.~\cite{Sullivan:1971kd, Burkardt:2012hk, Salamu:2014pka, Wang:2016ndh, Salamu:2018cny, Salamu:2019dok, Holtmann:1996be, Thomas:1983fh}.
A cutoff mass $\Lambda$ is invoked to regulate the ultraviolet divergence from the $k_\perp^2$ integration in the definition of $f_{\pi N}(\bar{x}_L)$, and is consequently a free parameter in our analysis, see Refs.~\cite{McKenney:2015xis, Barry:2018ort, Cao:2021aci}.

The structure function of the pion, evaluated at $x_\pi = x/\bar{x}_L$, is a convolution of the pion PDFs with the hard coefficients for DIS on a pion,
\begin{equation}
    F_2^\pi(x_\pi,Q^2) = \sum_i \int_{x_\pi}^1 \dd\xi\, f_i^\pi(x_\pi/\xi,\mu^2)\,
    \mathcal{C}_i^{\rm \tiny{DIS}}(\xi,\mu^2, Q^2).
    \label{eq.F2pi}
\end{equation}
Similar to Eq.~(\ref{eq.DY}), the range of the integral is from $x_\pi$ to 1.
However, unlike for the DY cross section, here the large logarithms near threshold do not contribute significantly to the perturbative coefficients.
For the treatment of the short-distance factorization coefficient, we therefore do not include threshold resummation, and calculate $\mathcal{C}^{\rm \tiny{DIS}}$ up to NLO in the strong coupling.

The measurements of the LN electroproduction were performed at HERA~\cite{H1:2010hym, ZEUS:2002gig}, and with focus on the low $x_\pi$ ($10^{-3} \lesssim x_\pi \lesssim 0.5$) and large $Q^2$ ($7<Q^2<10^3$~GeV$^2$) regions, provided access to pion PDFs across a wider range of kinematics than with DY experiments alone.
While the H1 experiment reported the LN structure function as in Eq.~(\ref{eq.F2LN}), in an effort to reduce systematic uncertainties the ZEUS experiment reported the ratio of the LN structure function to the inclusive structure function,
\begin{equation}
r(x,Q^2,x_L)
= \frac{\dd^3\sigma/\dd x\,\dd Q^2\,\dd x_L}{\dd^2\sigma/\dd x\,\dd Q^2}\, \Delta x_L,
\label{eq.r}
\end{equation}
where $\Delta x_L$ is the bin size in $x_L$ measured.
Cuts on the data are made for $\bar{x}_L$ to ensure the dominance of the one-pion exchange mechanism.
Phenomenologically, in Refs.~\cite{Barry:2018ort, McKenney:2015xis} the optimal limit was determined to be $\bar{x}_L < 0.2$, which leaves 58 and 50 data points for the H1 and ZEUS experiments, respectively.

%.....................................................................
\subsection{Lattice observables}

PDFs are formally defined as Fourier transforms of nonlocal matrix elements of quark and gluon fields evaluated on the light-front \cite{Collins:1989gx}.
The spacetime signature of the Euclidean lattice precludes direct calculation of matrix elements of lightlike separated quark and gluon fields, but does admit computation of the matrix elements of spacelike nonlocal operators. 
Unlike experimental cross sections,
lattice QCD calculated matrix elements of spacelike nonlocal operators are not physical observables, and their values depend on the choice of nonperturbative operator renormalization.
Consequently, perturbatively calculable coefficients for factorizing lattice calculated matrix elements will not only depend on the factorization scheme to relate to  PDFs,
but will also be sensitive to how the lattice calculated matrix elements are renormalized.
The matching coefficients take into account only the perturbative dependence of the renormalization scheme, and cannot account for nonperturbative effects in the renormalization of the spacelike nonlocal operators.

In this work, we focus on the lattice calculable matrix elements of spacelike separated quark-quark correlators linked by a straight Wilson line and CC correlators. 
In a manner similar to experimental cross sections, the renormalized matrix elements of these two types of correlators can be factorized into PDFs with perturbative matching coefficients in the SDF framework~\cite{Ma:2014jla, Ma:2017pxb}.
The resulted correlation functions provide complementary information on PDFs, however, the results are subjected to different systematic effects. 
The quark-quark correlator with a Wilson line has a power ultraviolet divergence, which must be removed nonperturbatively. 
Several nonperturbative renormalization schemes have been proposed in the literature~\cite{Constantinou:2017sej, Alexandrou:2017huk, Chen:2017mzz, Stewart:2017tvs, Orginos:2017kos, Braun:2018brg, Li:2020xml}.
In this paper, we use data from lattice QCD calculations with the renormalization defined in the reduced pseudo-PDF approach~\cite{Radyushkin:2017cyf}, which removes ultraviolet divergences through construction of suitable ratios of matrix elements.

The nonperturbative renormalization of CC operators is straightforward. 
In addition, different choices of currents provide additional sensitivities to different combination of PDFs, such as the direct access to the difference of quark and antiquark distributions~\cite{Sufian:2020vzb}.
CC~correlator calculations require lattice computations of four-point functions, instead of the three-point functions in the case of the quark-quark correlator with a Wilson line, making the task more computationally intensive.
Comparison of the data obtained from correlation functions of these two complementary correlators may provide a way to estimate systematic uncertainties associated with lattice calculations relevant to hadron structure.

% . . . . . . . . . . . . . . . . . . . . . . . . . . . . . . . . . . . . 
\subsubsection{Reduced Ioffe time pseudo-distributions}

The pion PDFs can be inferred from the Lorentz-invariant Ioffe time pseudo-distribution \mbox{(pseudo-ITD)}, $\mathcal{M}$, which is defined as
\begin{equation}
\mathcal{M}(\nu,z^2) 
= \frac{1}{2p^0}\,
  \langle p |\, \overline{\psi}(0)\gamma^0\, \mathcal{W}(z;0)\, \psi(z)\, |p \rangle,
    \label{eq.M}
\end{equation}
where $\mathcal{W}(z;0)$ is a straight gauge link of length $z$ in the fundamental representation, \mbox{$z=(0;0,0,z_3)$} is the displacement of the bare quark field, $\psi$, and $|p\rangle$ is the pion state with momentum $p$.
In the continuum limit, the pseudo-ITD is a function of only Lorentz invariant quantities, namely, the Ioffe time $\nu = p\cdot z$ \cite{Ioffe:1969kf} and the square of the separation.

The data used in this work are presented as a {\it reduced} pseudo-ITD (Rp-ITD) \cite{Joo:2019bzr},
\begin{equation}
    \mathfrak{M}(\nu,z^2)=\frac{\mathcal{M}(\nu,z^2)}{\mathcal{M}(0,z^2)},
    \label{eq.RpITD}
\end{equation}
which is the ratio of the function with respect to the $\bm{p} = 0$ distribution.
The ultraviolet divergences that are associated with the gauge link when $z$ is spacelike factorize multiplicatively and contain no dependence on $\nu$ \cite{Radyushkin:2018cvn}, and by taking the ratio of the pseudo-ITDs at $\nu$ to that at $\nu=0$ these divergences cancel.
This ratio is renormalization group invariant, so the factorization will not have the scheme dependence that other renormalization approaches would have.
Additionally, statistical uncertainties and a number of systematic uncertainties cancel \cite{Orginos:2017kos, Radyushkin:2017cyf}, resulting in smaller errors on the Rp-ITD.

To extract the PDFs of the pion from the Rp-ITD, we expand the Rp-ITD using an operator product expansion \cite{Karpie:2018zaz} in terms of local, nonperturbative, renormalized matrix elements and perturbatively calculable Wilson coefficients \cite{Radyushkin:2018cvn, Zhang:2018ggy, Izubuchi:2018srq}.
The factorization of $\mathfrak{M}\left(\nu,z^2\right)$ involves these coefficients and the light-cone Ioffe time distribution, whose Fourier transform with respect to $\nu$ gives the parton density $f_q(x)$ for flavor $q$.
The integration range of $x$ in the factorization formula is from $x=-1$ to 1, and from cross symmetry one has $f_q(x)=q(x)$ for positive $x$, and $f_q(x)=-\bar{q}(-x)$ for negative $x$, where $|x|$ corresponds to the momentum fraction of the quark and antiquark, respectively.
To obtain a meaningful description of the valence quark distribution in the pion, $q_v(x)=q(x)-\bar{q}(x)$, we take the real component of the factorized Rp-ITD $\mathfrak{M}(\nu,z^2)$ \cite{Joo:2019bzr}.
The imaginary component of $\mathfrak{M}(\nu,z^2)$ computed in these datasets cannot be expressed in terms of the physical pion sea quark distribution, because the calculation does not at present include the disconnected contributions.

The Rp-ITD takes the form of a convolution of the PDF with a hard coefficient function, to which correction terms are added to account for systematic uncertainties.
The real part of $\mathfrak{M}$ can be expressed as
\begin{eqnarray}
{\rm Re}~\mathfrak{M}(\nu,z^2)
&=& \int_0^1 \dd{x}\, q_v(x,\muL)\, 
    \mathcal{C}^{\rm Rp\mbox{-}ITD} \left(x\nu,z^2,\muL\right)
\notag\\
&+&\, z^2 B_1(\nu)\,
 +\, \frac{a}{|z|} P_1(\nu)\,
 +\, e^{-m_\pi (L-z)}F_1(\nu)\,
 +\, \ldots,
\label{eq.ReRpITDmatching}
\end{eqnarray}
where $q_v \equiv u-\bar{u}=\bar{d}-d$ in $\pi^+$, and the $\mathcal{C}^{\rm Rp\mbox{-}ITD}$ is the cosine transformation of the coefficient function, as discussed in Refs.~\cite{Karpie:2018zaz, Radyushkin:2018cvn, Zhang:2018ggy, Izubuchi:2018srq}.
The effects from threshold resummation on these types of matching coefficients for lattice observables exhibiting integrations from $x=0$ to 1 were shown to be small \cite{Gao:2021hxl}, and in this analysis we therefore keep the short distance coefficient $\mathcal{C}^{\rm Rp\mbox{-}ITD}$ to NLO.
In practice, we apply a truncation on the Taylor expansion of the cosine, expressing Eq.~\eqref{eq.ReRpITDmatching} in terms of the moments of the valence quark distribution and moments of the coefficient function $\mathcal{C}^{\rm \tiny{Rp-ITD}}$, which are calculable analytically \cite{Karpie:2018zaz}.
In Eq.~\eqref{eq.ReRpITDmatching}, $\muL$ refers to the scale at which the PDF and coefficient function are specified; more details on the choice of this scale appear in Sec.~\ref{ssec.scale}.

Beyond the leading twist term, the systematic correction terms in Eq.~(\ref{eq.ReRpITDmatching}) describe corrections relating to higher twist [$z^2 B_1(\nu)$], lattice spacing [$(a/|z|) P_1(\nu)$], and finite volume [$e^{-m_\pi (L-z)} F_1(\nu)$] effects, where $m_\pi$ is the pion mass characterizing the lattice ensemble, and $L$ is the spatial extent of the lattice.
These can be understood as matrix elements arising from higher twist operators in the OPE, Symanzik improvement of the lattice operators, and continuum finite volume calculations, respectively.
While these terms would generally have both $\nu$ and $z^2$ dependence, in practice we neglect the $z^2$ dependence in these terms since this is further suppressed by $\alpha_s$ or $(z \Lambda_{\rm QCD})^2$. 
There also exists a possibility for inclusion of target mass corrections proportional to powers of $m_\pi^2 z^2$. These effects appear from the trace terms in the OPE, just as when analyzing experimental cross-sections. In Ref.~\cite{Radyushkin:2017ffo}, models have been used to demonstrate that target mass effects will be quite small even when they would naively appear of $O(1)$.
The ellipsis in Eq.~\eqref{eq.ReRpITDmatching} represents higher order effects and other systematic uncertainties, such as $\mathcal{O}(a\Lambda_{\rm QCD})$ discretization and pion mass corrections, which cannot be probed given the constraints of the currently available data.
Additional systematic effects may also include excited state contamination in the lattice data, which for the pion are less important than in the proton case \cite{Lepage:1989hd}, and perturbative truncation errors.

Each of the functional forms of the accessible systematic correction terms in Eq.~\eqref{eq.ReRpITDmatching} is parametrized by a series of $\nu$-dependent functions, as in Ref.~\cite{Karpie:2021pap}, given by
\begin{subequations}
    \label{eq.BPF1}
\begin{eqnarray}
B_1(\nu) &=& \sum_n \sigma_{0,n}(\nu)\, b_{n},  \\
    \label{eq.B1}
P_1(\nu) &=& \sum_n \sigma_{0,n}(\nu)\, p_n,    \\
    \label{eq.P1}
F_1(\nu) &=& \sum_n\sigma_{0,n}(\nu)\, f_n,
    \label{eq.F1}
\end{eqnarray}
\end{subequations}
where $\sigma_{0,n}(\nu)$ is defined as
\begin{equation}
\sigma_{0,n}(\nu) 
= \int_0^1 \dd{x} \cos(\nu x)\, x^a (1-x)^b\, J^{(a,b)}_n(x).
    \label{eq.sigma0}
\end{equation}
Here, $J^{(a,b)}_n(x)$ is a transformed Jacobi polynomial existing in the range $0\leq x \leq 1$. 
Since the set of all $J^{(a,b)}_n(x)$ forms a complete basis of functions for all $a,b > -1$, an infinite series of $\sigma_{0,n}$ could reproduce any function of Ioffe time for any $a,b >-1$.
In this analysis, we include $a$ and $b$ in the set of fit parameters.
By construction, $\mathfrak{M}(0,z^2)=1$, and the leading twist contribution satisfies this condition through the valence quark number sum rule.
At $\nu=0$ the systematic corrections must therefore vanish.
For this reason, we do not include the $n=0$ term in the series in Eqs.~\eqref{eq.BPF1}, since these are not guaranteed to vanish at $(n,\nu)=(0,0)$.
Instead, we begin the expansion at $n=1$ and take up to the $n=2$ term, since both terms vanish at $\nu=0$.

\begin{table}
\caption{Parameters for the gauge ensembles used in this analysis: lattice spacing ($a$), pion mass ($m_\pi$), inverse gauge coupling ($\beta$), spatial ($L$) and temporal ($T$) sizes, and the smallest momentum in each ensemble ($p_1$) \cite{Joo:2019bzr, Sufian:2020vzb}. Note that while in Ref.~\cite{Joo:2019bzr} the ID names are {\tt a127m415} and {\tt a127m415L} for $m_\pi\approx 415~{\rm MeV}$, the entries in this table for {\tt a127m413} and {\tt a127m413L} represent the same lattice ensembles that were used in \cite{Joo:2019bzr}.\\}
\begin{tabular}{ l | c c | c c | c }
\hline
~~~~~~ID~~~~~~~ & ~~~$a$~(fm)~~~ & ~~$m_\pi$~(MeV)~~ & ~~~$\beta~$~~~ & ~~$L^3 \times T$~~~ & ~~~$p_1$~(MeV)~~~\\
\hline
~~{\tt a127m413} & 0.127(2) & 413(4) & 6.1 & $24^3 \times 64$ & 406(6) \\
~~{\tt a127m413L}~ & 0.127(2) & 413(5) & 6.1 & $32^3 \times 96$ & 305(5) \\
\hline
~~{\tt a94m358} & 0.094(1) & 358(3) & 6.3 & $32^3 \times 64$ & 411(4) \\
~~{\tt a94m278} & 0.094(1) & 278(4) & 6.3 & $32^3 \times 64$ & 411(4) \\
\hline
\end{tabular}
\label{t.lat}
\end{table}

We use the available Rp-ITD data from the Jefferson Lab HadStruc group \cite{Joo:2019bzr}, calculated on the ensembles labeled $\tt a127m413$ and $\tt a127m413L$, whose parameters are given in Table~\ref{t.lat}.
The two datasets each have the same lattice spacing $a \approx 0.127~{\rm fm}$ and pion mass $m_\pi \approx 413~{\rm MeV}$, but different lattice volumes, with the larger volume dataset $\tt a127m413L$ having $L=32a$ and the smaller volume dataset $\tt a127m413$ having $L=24a$.
Each dataset includes the momentum values of $p_i=i \times (2\pi/La)$, where $i \in \{1,2,3\}$, and each featuring discrete separations $z$ up to $6a$, with $\tt a127m413L$ including $z$ up to $8a$.
The signal-to-noise ratio of the lattice points decays exponentially with momentum, making the calculation of the $p_3$ points in the $\tt a127m413$ ensemble incompatible with statistical fluctuations.
For this reason, we exclude these points.
We also cut the $z\leq2a$ data points to avoid discretization errors and contact terms arising from the chosen discretization of the quark action.

% . . . . . . . . . . . . . . . . . . . . . . . . . . . . . . . . . . . . 
\subsubsection{Current-current correlators}

CC correlators provide another class of lattice calculable quantities that can be factorized into PDFs convoluted with perturbative matching kernels~\cite{Ma:2017pxb}.
The relevant available data~\cite{Sufian:2020vzb} were obtained from a pair of vector and axial-vector currents, which are written in the form
\begin{equation}
\Sigma^{\alpha\beta}_{VA}(z,p) 
= z^4 Z_V Z_A\,
\langle p |\,
    [\bar{\psi}\gamma^\alpha \psi](z)\,
    [\bar{\psi}\gamma^\beta \gamma^5 \psi](0)\,
| p\, \rangle\, 
+\, \big( V \leftrightarrow A \big),
    \label{eq.CCcorrelator}
\end{equation}
where $Z_V$ and $Z_A$ are the renormalization constants of the local lattice vector and axial-vector currents, respectively.
Through time-reversal and parity invariance, the CC correlator in Eq.~\eqref{eq.CCcorrelator} is antisymmetric in the Lorentz indices $\{\alpha,\beta\}$.
The CC correlator can then be expressed as two dimensionless pseudo-structure functions, $T_1$ and $T_2$, which are functions of the Lorentz invariants $\nu$ and $z^2$ \cite{Sufian:2019bol}.
Choosing the momentum $p=(p^0;0,0,p^3)$
with the Lorentz indices $\alpha=1$, $\beta=2$, we isolate the function $T_1$, which can be related to the light-cone valence quark PDF by
\begin{equation}
T_1(\nu,z^2) = \int_0^1 \dd x\, q_v(x,\muL)\, \mathcal{C}^{\rm \tiny{CC}}(x\nu,z^2,\muL)
+ z^2 B_1(\nu) + a R_1(\nu) + \ldots,
    \label{eq.T1matching}
\end{equation}
where $\mathcal{C}^{\rm \tiny{CC}}$ is the matching coefficient, analytically computable through perturbative QCD \cite{Ma:2017pxb, Sufian:2019bol, Sufian:2020vzb, Li:2020xml}, which is taken up to NLO in this work.
As is the case with the Rp-ITD, the threshold resummation for $\mathcal{C}^{\rm \tiny{CC}}$ is less important over the whole range of $0 \leq x \leq 1$ than for experimental observables.
The leading twist term of Eq.~(\ref{eq.T1matching}) is very similar to the leading twist term in Eq.~(\ref{eq.ReRpITDmatching}), where $\mathcal{C}^{\rm \tiny{CC}}$ is an implied integral quantity involving a cosine transformation \cite{Sufian:2020vzb}.

The systematic effects added to the leading twist term in Eq.~\eqref{eq.T1matching} are the power correction $z^2 B_1(\nu)$ and discretization correction $a R_1(\nu)$, with the ellipsis representing other potential systematic correction terms.
Attempts were made to fit systematic corrections such as lattice spacing, pion mass, and finite volume, but because the data have such large statistical uncertainties, it was difficult to separate the systematic effects beyond what is included in Eq.~\eqref{eq.T1matching}.
We parametrize function $B_1(\nu)$ as in Eq.~\eqref{eq.B1}, and the discretization term $R_1(\nu)$ is similarly parameterized as
\begin{equation}
R_1(\nu) = \sum_n\sigma_{0,n}(\nu)\, r_n,
    \label{eq.R1}
\end{equation}
where $\sigma_{0,n}$ is given in Eq.~\eqref{eq.sigma0}.
Unlike in the Rp-ITD case, here we expand Eqs.~\eqref{eq.B1} and \eqref{eq.R1} in $n$ from $0$ to $2$ because the quantity $T_1(\nu,z^2)$ does not have a specified normalization given by the leading twist term at $\nu=0$.

Four datasets were included in our analysis of the HadStruc data~\cite{Sufian:2020vzb}, with the parameters given in Table~\ref{t.lat}.
We refer to the datasets as 
    \mbox{\{{\tt a127m413}, $\tt a127m413L$, $\tt a94m358$, $\tt a94m278$\}},
which have lattice spacings of 
    \{0.127, 0.127, 0.094, 0.094\}~fm
and pion masses of
    \{413, 413, 358, 278\}~MeV,
respectively.
Each ensemble has a total lattice size of $L=32a$, with the exception of the $\tt a127m413$ set, which has $L=24a$.
Each dataset spans the range $2a \leq z \leq 4a$, and the $\tt a94m358$ and $\tt a94m278$ ensembles further include $z=5a$ and $6a$.
The momentum of the generated lattice data include $p_i=i \times (2\pi/La)$, where $i \in \{1,2,3,4\}$.
No kinematic cuts were made on the CC observables.

%.....................................................................
\subsection{Scale setting}
\label{ssec.scale}

In typical high energy scattering experiments, such as DY lepton-pair production or deep-inelastic LN electroproduction, the invariant mass of the virtual photon $Q\equiv\sqrt{Q^2}$ is much greater than any nonperturbative scale, rendering the power correction terms in factorized cross sections small.
The renormalization and factorization scales relating the PDFs and hard coefficients to the experimental cross sections are generally set to this probing scale, $\mu=Q$, where asymptotic freedom is exploited and $\log(\mu^2/Q^2)$-type logarithms appearing in the perturbative expansions are suppressed.

The uncertainty in renormalized correlation functions of lattice calculated correlators is independent of how they are factorized into PDFs and the corresponding choice of renormalization and factorization scale, although it impacts the calculation of the perturbative matching coefficients.
In the Rp-ITD case, the renormalization constants of lattice calculated ITDs, which have been calculated in Ref.~\cite{Yoon:2016jzj}, are multiplicative and cancel in the ratio in Eq.~(\ref{eq.RpITD}).
For CC observables, the renormalization constants are explicit in Eq.~(\ref{eq.CCcorrelator}), given by $Z_V$ and $Z_A$.
To factorize these renormalized lattice calculated correlators into PDFs convoluted with hard coefficients we need to specify the factorization and renormalization scales, 
akin to the scales that appear in the treatment of experimental observables.
We choose to equate these factorization and renormalization scales, denoted by $\muL$ in Eqs.~\eqref{eq.ReRpITDmatching} and \eqref{eq.T1matching}.
It is convenient to choose this scale to be proportional to the inverse of the separation between fields or currents in order to eliminate large logarithms \cite{Radyushkin:2017cyf}.

It was shown in Ref.~\cite{Joo:2019bzr} that the power corrections of $\mathcal{O}(z^2 \Lambda_{\rm QCD}^2)$ are much less than the leading twist component, and the lattice data are well behaved when $1/z < 1$~GeV.
We fix $\muL$ to be a constant across all kinematics, satisfying two requirements: {\bf (i)}~$\alpha_s(\muL)$ should not be too large, and {\bf (ii)}~the product of $\alpha_s(\muL)$ and the logarithm appearing in the perturbatively calculable matching coefficients $\sim \log(z^2 \muLsq)$ is not too large.
We investigate these criteria for three scales: $\muL=m_c=1.27~{\rm GeV}$, $\muL=2~{\rm GeV}$, and $\muL=4~{\rm GeV}$.
The first requirement is satisfied by all three choices, as the scales provide $\alpha_s\simeq 0.37$, $0.30$, and $0.23$, respectively, in the $\overline{\rm MS}$ renormalization scheme at NLO.

\begin{figure}
    \centering
    \includegraphics[width=0.8\textwidth]{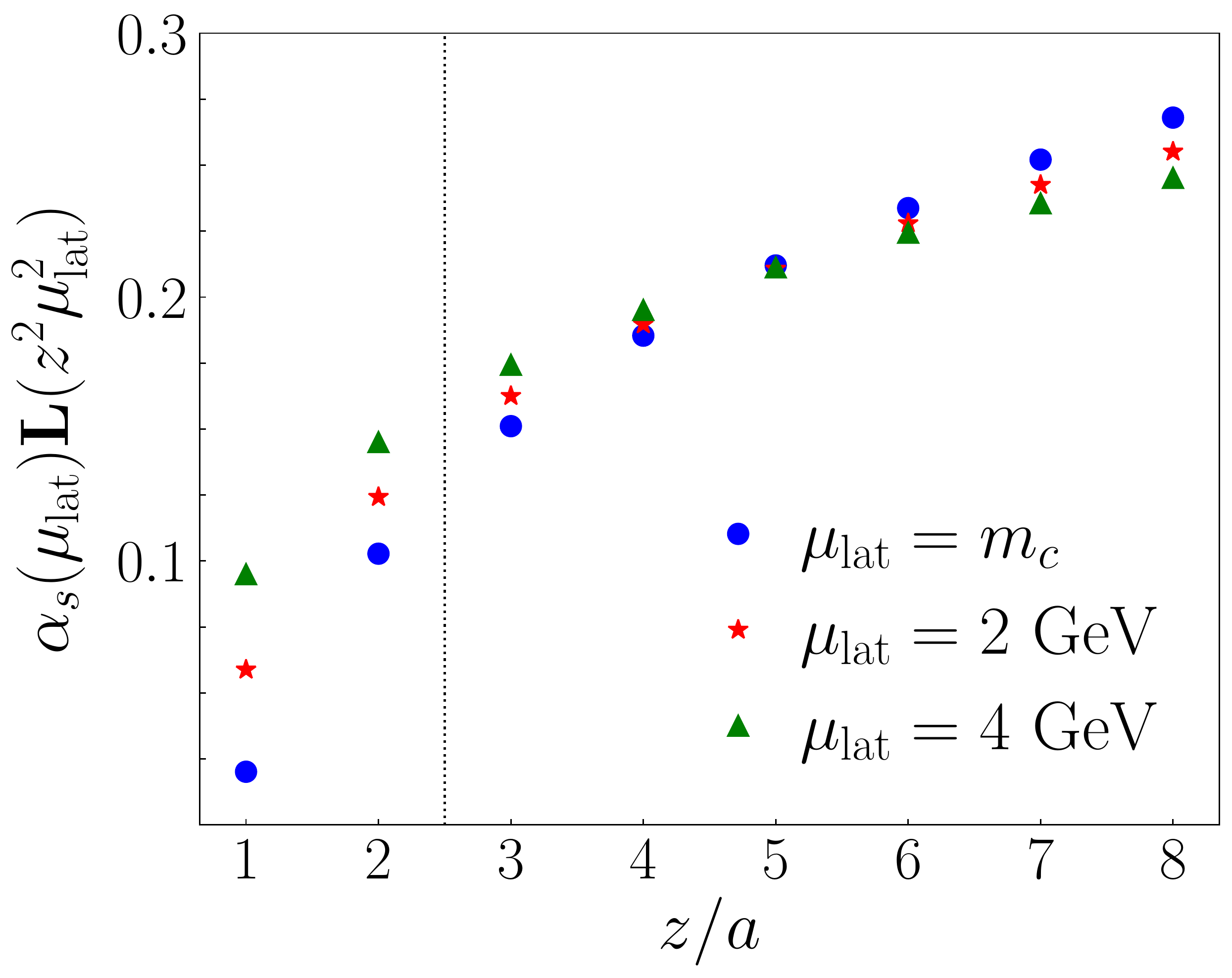}
    \caption{Typical factor of $\mathcal{O}(\alpha_s)$ in the matching coefficients with explicit dependence on $z$ and $\muL$, where 
    ${\bf L}(z^2\muLsq) \equiv \log(\frac14 z^2\muLsq e^{2\gamma_E + 1})/2\pi$.
    Shown are the terms evaluated using 3 fixed scales of $\muL=m_c$ (blue circles), $\muL=2~{\rm GeV}$ (red stars), and $\muL=4~{\rm GeV}$ (green triangles) as a function of $z$ in units of lattice spacing $a=0.127~{\rm fm}$.
    The dotted line indicates the lower limit of $z/a$ used in the analysis.}
    \label{f.aSlog}
\end{figure}

The interplay between $\alpha_s(\muL)$ and $\log(z^2 \muLsq)$ in criterion {\bf (ii)} is illustrated in Fig.~\ref{f.aSlog}, where the term $\alpha_s(\muL){\bf L}(z^2\muLsq)$ is shown as a function of $z$ in units of $a=0.127$~fm, with
    ${\bf L}(z^2\muLsq) \equiv \log(\tfrac14 z^2\muLsq e^{2\gamma_E+1})/2\pi$.
The product of $\alpha_s$ and the logarithm ${\bf L}$ is well below unity, indicating that perturbation theory can be trusted at each of the choices of constant $\muL$ for the kinematics provided.
While the results in Fig.~\ref{f.aSlog} are shown at $a=0.127$~fm, similar types of logarithms that appear in $\mathcal{C}^{\rm \tiny{CC}}$ for the CC correlators in the form     ${\bf L}(z^2\mu^2) = \log(\frac14 z^2\mu^2 e^{2\gamma_E})/2\pi$
for all values of $a$ are also well below unity, with the largest value of $\alpha_s(\muL){\bf L}(z^2\muLsq) = 0.17$.
In the remainder of this paper, we choose the scale $\muL=2~{\rm GeV}$, unless otherwise noted.

%.....................................................................
\subsection{Methodology}

We note that each of the expressions in Eqs.~\eqref{eq.DY}, \eqref{eq.F2pi}, \eqref{eq.RpITD}, and \eqref{eq.T1matching} are in the form of a convolution of a PDF (or two PDFs in the DY case) and a hard coefficient function, $\mathcal{C}$.
The observables in these equations do not provide a direct method for solving for the PDFs, whose analytic form is not known, on a point-by-point basis, and therefore the extraction of PDFs from these equations constitutes an inverse problem.

Following the standard approach used in the literature, we parametrize the valence quark distribution $q_v$, sea quark distribution $q_s$, and the gluon distribution $g$ at the initial scale $\mu_0$ through the general template function,
\begin{equation}
f(x,\mu_0^2) = \frac{N_f\, x^{\alpha_f} (1-x)^{\beta_f}(1+\gamma_f x^2)}{B(\alpha_f + 2, \beta_f + 1) + \gamma_f B(\alpha_f + 4,\beta_f + 1)},
    \label{eq.PDF}
\end{equation}
where $f\, (= q_v$, $q_s$ or $g$) labels the parton flavor.
The input scale is chosen to be the charm quark mass evaluated at the scale of the charm mass in the $\overline{\mathrm{MS}}$ scheme, $\mu_0=m_c(m_c)=1.27~{\rm GeV}$.
In the remainder of the text, we assume the scale and refer to $m_c(m_c)$ simply as~$m_c$.
We impose the valence quark number rule and the momentum sum rule, which fix the valence quark and sea quark normalizations, respectively.
Even though the lattice data are not sensitive to the sea quark and gluon distributions, they may have an effect through the momentum sum rule and interplay with the effects from the DY and LN experimental data.
In a previous analysis \cite{Cao:2021aci}, it was found that more flexible parametrizations did not change significantly the resulting PDFs and the agreement with the data.
Since the parameters $\gamma_s$ and $\gamma_g=0$ are not constrainable by the data, we set them both to zero.
However, we allow $\gamma_v$ for the valence quark PDF to be free.
In total, there are eight PDF shape parameters to be fitted, along with the $\Lambda$ cutoff parameter for the LN observables.
The scale dependence of the PDFs is determined by solving the DGLAP equations using the zero-mass flavor number scheme evolved up to next-to-leading logarithmic accuracy.

As in previous JAM global QCD analyses of pion PDFs~\cite{Barry:2018ort, Cao:2021aci, Barry:2021osv}, we employ Bayesian statistics to sample the posterior distribution according to
\begin{equation}
\mathcal{P}(\bm{a}|{\rm data}) 
\sim \mathcal{L}({\rm data}|\bm{a})\, \pi(\bm{a}),
\label{eq.Bayes}
\end{equation}
where $\pi({\bm a})$ is the prior distribution, which sets the boundaries of the fit parameters $\bm{a}$, ${\cal L}$ is the likelihood function,
\begin{equation}
\mathcal{L}({\rm data}|\bm{a}) 
= \exp \Big( \!-\frac12 \chi^2(\bm{a},{\rm data}) \Big),
\label{eq.likelihood}
\end{equation}
which is a Gaussian with the argument $\chi^2$.
In practice, maximizing the likelihood function is done by minimizing the $\chi^2$ function, which for a given experimental dataset $e$ is defined by
\begin{equation}
\chi^2_e(\bm{a},{\rm data})
= \sum_i 
  \bigg[
   \frac{d_i^{\, e} - \sum_k r^e_k\, \beta^{\, e}_{k,i} - t^e_i(\bm{a})/n_e}
        {\alpha^e_i}
  \bigg]^2
+ \left(\frac{1-n_e}{\delta n_e}\right)^2 + \sum_k \big( r_k^e \big)^2,
\label{eq.chi2expt}
\end{equation}
where the sum is over $i$ data points up to $N_{\rm dat}^e$, with $k$ types of correlated uncertainties.
The data points are represented by $d_i^{\, e}$, while the parameter-dependent theory, $t_i^e(\bm{a})$, is divided by the fitted normalization parameter $n_e$ for each experiment and added with the correlated shift, which includes the nuisance parameters $r_k^e$ and point-to-point correlated uncertainties $\beta_{k,i}^{\, e}$.
In the second term the $\delta n_e$ is the overall normalization uncertainty reported by the experiment.
The last two terms in Eq.~(\ref{eq.chi2expt}) represent penalties such that the fit is disfavored if $n_e$ is far from 1, or if the amount of shift needed on the theory, $r_k^e$, is large.

Lattice data points are correlated within the same ensemble $\lambda$, so the point-to-point statistical uncertainties are not treated in the same way as the uncertainties on the experimental data.
Instead, we use the covariance matrix $V$ and minimize
\begin{equation}
\chi^2_{\rm \lambda}(\bm{a},{\rm data}) 
= \left( \bm{D}^\lambda - \bm{T}^\lambda(\bm{a})\right)^T 
  V_\lambda^{-1} 
  \left( \bm{D}^\lambda - \bm{T}^\lambda(\bm{a})\right),
    \label{eq.chi2lat}
\end{equation}
where $\bm{D}^\lambda$ is the vector of data points of length $N_{\rm dat}^\lambda$, and $\bm{T}^\lambda$ is the set of theory predictions corresponding to the data points.
The covariance matrix $V$ follows from the standard definition,
    $V_{ij}={\rm E}
    \big[
    (\mathcal{O}_i-{\rm E}[\mathcal{O}_i])
    (\mathcal{O}_j-{\rm E}[\mathcal{O}_j])
    \big]$,
where $\rm E$ is the expectation value defined in Eq.~\eqref{eq.EV} below.
The total $\chi^2$, as in Eq.~\eqref{eq.likelihood}, that governs the parameter determination in each minimization is 
    $\chi^2 = \sum_e \chi^2_e + \sum_\lambda \chi^2_\lambda$.

In our Monte Carlo analysis the $\chi^2$ minimizations are executed $N_{\rm rep}$ number of times.
Through data resampling for each minimization, Gaussian noise is added to the central value of the data point with $1\sigma$ uncertainty width.
For $\chi^2$ minimizations using Eq.~(\ref{eq.chi2lat}), we diagonalize the covariance matrix and add Gaussian noise in each eigendirection with width given by the magnitude of the eigenvalue.
In this way we populate statistics on the posterior distribution, and avoid local minima in parameter space.
The resulting parameter sets $\bm{a}_j$, where $j$ runs from 1 to the number of replicas, $N_{\rm rep}$, are used to obtain expectation values $\rm E$ and variances $\rm V$ of observables $\mathcal{O}$, defined as
\begin{subequations}
\begin{align}
    {\rm E}[\mathcal{O}] &= \frac{1}{N_{\rm rep}} \sum_j \mathcal{O}({\bm a}_j),
    \label{eq.EV}
    \\
    {\rm V}[\mathcal{O}] &= \frac{1}{N_{\rm rep}} \sum_j \left(\mathcal{O}({\bm a}_j) - {\rm E}[\mathcal{O}] \right)^2.
\end{align}
\end{subequations}
The agreement between data and theory is quantified using the reduced $\chi^2$, which is defined here as
\begin{subequations}
\begin{align}
\overline{\chi}^2_e 
&= \frac{1}{N_{\rm dat}^e}\sum_i 
\left[
\frac{d_i^{\, e} - {\rm E}[\sum_k r^e_k\, \beta^{\, e}_{k,i} + t^e_i(\bm{a})/n_e]}
  {\alpha^e_i}
\right]^2,
    \label{eq.rchi2e}
\\
\overline{\chi}^2_\lambda 
&= \frac{1}{N_{\rm dat}^\lambda}
\left( 
\bm{D}^\lambda - {\rm E}[\bm{T}^\lambda(\bm{a})]\right)^T V_\lambda^{-1} \left( \bm{D}^\lambda - {\rm E}[\bm{T}^\lambda(\bm{a})]\right)
    \label{eq.rchi2l},
\end{align}
\end{subequations}
or, generally, $\overline{\chi}^2$.
For our analysis, an acceptable $\overline{\chi}^2$ value for each dataset is around 1.
The total reported reduced $\chi^2$ is given by
    $\overline{\chi}^2_{\rm tot}
    = \big( \sum_e \overline{\chi}^2_e N_{\rm dat}^e 
          + \sum_\lambda \overline{\chi}^2_\lambda N_{\rm dat}^\lambda 
      \big) / N_{\rm dat}^{\rm tot}$,
where $N_{\rm dat}^{\rm tot}$ is the grand total number of all data points.

%%%%%%%%%%%%%%%%%%%%%%%%%%%%%%%%%%%%%%%%%%%%%%%%%%%%%%%%%%%%%%%%%%%%%%
\section{QCD Analysis Results}
\label{s.results}

While most previous analyses of pion PDFs performed fits to DY data with an NLO expansion for the hard coefficients, several more recent analyses utilized the NLO+NLL approximation with threshold resummation.
Both of these approaches give good descriptions of the DY and LN data \cite{Barry:2021osv}, so that it is difficult to assess the applicability of factorization theorems if only one observable is present, since changes in the hard factors can generally be accommodated by changes in the PDFs.
For this reason we present results with NLO calculations for the hard coefficients (referred to as ``NLO''), as well as an NLO+NLL calculation for DY coupled with NLO calculations for the remaining hard coefficients (referred to as ``NLO+NLL$_{\rm DY}$'').
Including lattice data in the mix may provide observables that allow us to test the degree of universality for PDFs across experimental and lattice data when using either fixed order or threshold resummation hard coefficients in DY.
Lattice data, as we shall see, have sizable systematic effects that currently prevent us reaching definite conclusions about the applicability of threshold correction for DY at large $x$.

%.....................................................................
\subsection{Analysis with reduced Ioffe time pseudo-distributions}
\label{s.RpITD}

As baseline fits for our analysis, we consider first as ``Scenario A'' the analysis of DY and LN data determined using the NLO or NLO+NLL$_{\rm DY}$ methods for the DY cross sections~\cite{Barry:2018ort, Barry:2021osv}, without inclusion of lattice data.
A good overall fit to the DY and LN data is obtained, similar to that in the recent JAM analysis~\cite{Barry:2021osv}.
As illustrated in Table~\ref{t.chi2_RpITD}, which summarizes the goodness of fit for each scenario, the number of points in each dataset and the corresponding reduced $\chi^2$, a total reduced $\overline{\chi}^2$ of $\approx 0.8$ is found for both the NLO and NLO+NLL$_{\rm DY}$ baseline fits to the experimental data.

\begin{table}
\centering
\caption{Summary of results from global fits to the DY cross sections, LN electroproduction data from HERA, and the RpITD data, including the number of data points $N_{\rm dat}$ and the $\overline{\chi}^2$ values for the NLO and NLO+NLL$_{\rm DY}$ methods. The Rp-ITD data were fitted at the scale $\muL=2~{\rm GeV}$. Scenario~A represents the fit to only experimental data, Scenario~B is fitting both experimental and lattice data with only the leading twist term, and Scenario~C (in boldface) is the full fit including systematic corrections.\\}
\begin{tabular}{llc|c|c|c}
\hline
&&& ~~~Scenario~A~~~
& ~~~Scenario~B~~~
& ~~~{\bf Scenario~C}~~~ \\
&&& ~~NLO~~~+$\rm NLL_{DY}$~~
& ~~NLO~~~+$\rm NLL_{DY}$~~
& ~~NLO~~~+$\rm NLL_{DY}$~~ \\

~Process~~ & ~Experiment~~  
        & ~~~$N_{\rm dat}$~~~  
        & ~~~$\overline{\chi}^2$  
        & ~~~$\overline{\chi}^2$  
        & ~~~$\overline{\chi}^2$\\
\hline
{\bf ~DY}& ~E615                    & 61 
                                    & 0.84~~~~~~0.82
                                    & 0.83~~~~~~0.82
                                    & $\bm{0.84~~~~~~0.82}$\\
         & ~NA10~\mbox{\scriptsize (194~GeV)}
                                    & 36
                                    & 0.53~~~~~~0.53
                                    & 0.52~~~~~~0.54
                                    & $\bm{0.51~~~~~~0.53}$\\
         & ~NA10~\mbox{\scriptsize (286~GeV)} 
                                    & 20
                                    & 0.80~~~~~~0.81
                                    & 0.78~~~~~~0.79
                                    & $\bm{0.74~~~~~~0.81}$\\
{\bf ~LN}& ~H1                      & 58 
                                    & 0.36~~~~~~0.35
                                    & 0.39~~~~~~0.39
                                    & $\bm{0.38~~~~~~0.37}$\\
         & ~ZEUS                    & 50
                                    & 1.56~~~~~~1.48
                                    & 1.62~~~~~~1.69
                                    & $\bm{1.59~~~~~~1.62}$\\ 
\hline
{\bf ~Rp-ITD} & ~\tt a127m413L          & 18
                                    & --~~~~~~--
                                    & 1.04~~~~~~1.06
                                    & $\bm{1.05~~~~~~1.04}$\\
              & ~\tt a127m413           & 8
                                    & --~~~~~~--
                                    & 1.98~~~~~~2.63
                                    & $\bm{1.00~~~~~~1.18}$\\ \hline
~\textbf{Total} &                   &{\bf 251}
                                    &{\bf 0.82}~~~~~~{\bf 0.80}
                                    &{\bf 0.89}~~~~~~{\bf 0.92}
                                    &{\bf 0.85}~~~~~~{\bf 0.86}   \\ \hline \\
\end{tabular}
\label{t.chi2_RpITD}
\end{table}

In a first attempt to fit the Rp-ITD data simultaneously with the experimental data, in ``Scenario B'' we include the lattice data in a simplified way, excluding the systematic effects and only fitting to the leading twist terms in Eq.~\eqref{eq.ReRpITDmatching}.
While the DY and LN data were described almost the same as the fits to only experimental data, we found the fit to the smaller lattice volume data, {\tt a127m413}, was unacceptably large, with a $\overline{\chi}^2$ of 1.98 and 2.63 for the NLO and NLO+NLL$_{\rm DY}$ methods, respectively.
This lack of agreement with the data is an indication that the leading twist convolution term alone is not sufficient to describe the available Rp-ITD data.
Consequently, we consider a further ``Scenario C,'' which includes all of the available types of systematic terms as shown in Eq.~\eqref{eq.ReRpITDmatching}.

With the inclusion of the systematic corrections in ``Scenario C,'' the resulting $\overline{\chi}^2$ are acceptable for all datasets, with a global $\overline{\chi}^2 \approx 0.85$.
Importantly, the $\overline{\chi}_e^2$ values do not change much among the scenarios, indicating compatibility of the experimental and lattice data.
Interestingly, the agreement of the results with the large lattice volume data, {\tt a127m413L}, does not change with the inclusion of systematic corrections, and the effect is largely seen in the smaller lattice volume ensemble, {\tt a127m413}.
By performing Monte Carlo inference of all PDF shape parameters and lattice systematic parameters, in the following we investigate the relationship between the PDFs and the systematic uncertainties, including their overall contributions.

\begin{table}[b]
\centering
\caption{Summary of $Z$-sigma levels for the $\chi^2$ distributions for each of the data sets.
The absolute $\sigma$-value by which the peak of the resulting Monte Carlo $\chi^2$ distribution is less than ($^-$) or greater than ($^+$) the expected $\chi^2$ distribution is shown in each entry.
\\}
\begin{tabular}{ll|c|c}
\hline
&& ~~NLO~~~
& ~~NLO+$\rm NLL_{DY}$~~\\
\hline

~Process~~ & ~~~Experiment~~~~
& ~~~$Z$-sigma level~~~~~  
& ~~~$Z$-sigma level~~~~~\\
\hline
{\bf ~DY}&~~~E615~~~ & 0.75$^-$ & 0.79$^-$\\
& ~~~NA10~\mbox{\scriptsize (194~GeV)}~~~~ & 2.32$^-$ & 2.26$^-$ \\
& ~~~NA10~\mbox{\scriptsize (286~GeV)}~~~~ & 0.82$^-$ & 0.78$^-$ \\
{\bf ~LN}&~~~H1~~~~ & 4.15$^-$ & 4.22$^-$\\
& ~~~ZEUS~~~~ & 2.72$^+$ & 2.73$^+$ \\ 
\hline
{\bf ~Rp-ITD}&~~~\tt a127m413L~~~~~ & 0.30$^+$ & 0.35$^+$ \\
&~~~\tt a127m413~~~~~ & 0.40$^+$ & 0.60$^+$ \\ 
\hline
\end{tabular}
\label{t.Zlevel}
\end{table}

To assess the significance of the $\bar{\chi}^2$ values in Table~\ref{t.chi2_RpITD}, we perform the $Z$-sigma level statistical test on each of the datasets analyzed in the global analysis.
In this test, the null hypothesis is the expected $\chi^2$ distribution assuming the number of degrees of freedom to be the number of points in the data set, and the resulting $\chi^2$ samples from the Monte Carlo (MC) analyses from ``Scenario C'' are the alternative hypotheses.
We compute the $p$-value using the peak of the resulting MC $\chi^2$ distributions, and the $Z$-sigma level is the inverse of the normal cumulative distribution function, $Z=\Phi^{-1}(p) \equiv \sqrt{2}~{\rm erf}^{-1}(2p-1)$.
These values are given in Table~\ref{t.Zlevel} for each of the data sets and represent the number of normal standard deviations from the expected $\chi^2$.
The ``$-$'' or ``$+$'' superscript on the values indicates their positions below or above the expected $\chi^2$ distribution's mean value, respectively.
A large negative value may imply a non-Gaussian $\chi^2$ function is needed to treat the systematic uncertainties.
Nevertheless, none of the values are outside of $5\sigma$, indicating a reasonable probability that our results are achieved with the given data.
Notably, the NLO and NLO+NLL$_{\rm DY}$ are not in tension for the Rp-ITD data.

\begin{figure}
    \centering
    \includegraphics[width=0.7\textwidth]{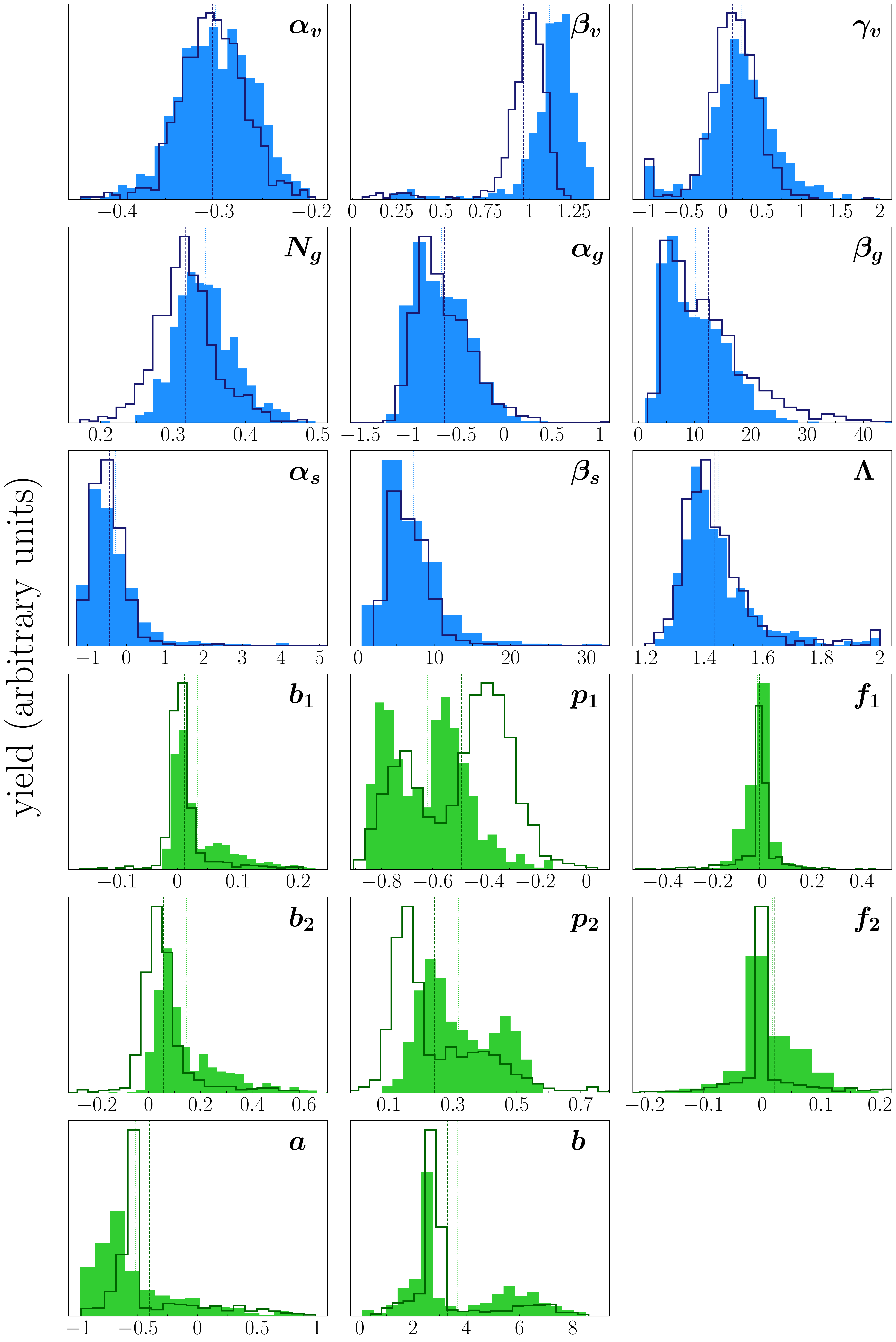}
    \caption{Distributions of the free parameters describing PDF shapes
        \mbox{\{$\alpha_v$, $\beta_v$, $\gamma_v$;
                $N_g$, $\alpha_g$, $\beta_g$;
                $\alpha_s$, $\beta_s$\}}
    and the ultraviolet cutoff mass $\Lambda$ in the $N \to \pi N$ splitting function (blue histograms), and the intrinsic lattice systematic parameters 
        \{$b_1$, $p_1$, $f_1$; $b_2$, $p_2$, $f_2$; $a$, $b$\}
    (green histograms) from the Monte Carlo fits, in arbitrary units.
    Both the NLO (outlined histograms) and NLO+NLL$_{\rm DY}$ (filled histograms) analyses are shown, with the vertical dotted lines giving the mean values of the parameters.}
    \label{f.parameters}
\end{figure}

The distributions of the free parameters from ``Scenario C'' are displayed in Fig.~\ref{f.parameters}, which shows the number of Monte Carlo samples with a best fit value in each bin, up to an arbitrary normalization, for both the NLO and NLO+NLL$_{\rm DY}$ methods.
The first three rows indicate the PDF shape parameters, as well as the LN cutoff parameter $\Lambda$, while the bottom two rows show the distribution of the systematic parameters $b_i$, $p_i$ and $f_i$ ($i=1,2$).
Note that the $\beta_v$ parameter shown in the first row is not the same as the $\beta_{\rm eff}$ that is used to describe the large-$x$ behavior of the valence quark distribution (see below).
In each panel a distinct peak is observed, indicating that the data prefer a certain value and assuring each parameter has settled in a global maximum in the likelihood.

The power correction parameters $b_1$ and $b_2$ both have generally positive values, with the $b_1$ parameter smaller and peaking slightly above zero, while the $b_2$ parameter peaks at around 0.1.
The lattice spacing error parameters $p_1$ and $p_2$ have opposite sign, with the $p_1$ parameter roughly twice as large as $p_2$ in magnitude.
The finite volume parameters are seen to have the most narrow distributions and have the smallest absolute values.
Both $f_1$ and $f_2$ are close to zero, with $f_1$ tending slightly negative and $f_2$ slightly positive.
We discuss the systematic corrections in more detail in Sec.~\ref{ssec.syst}.

\begin{figure}
    \centering
    \includegraphics[width=0.9\textwidth]{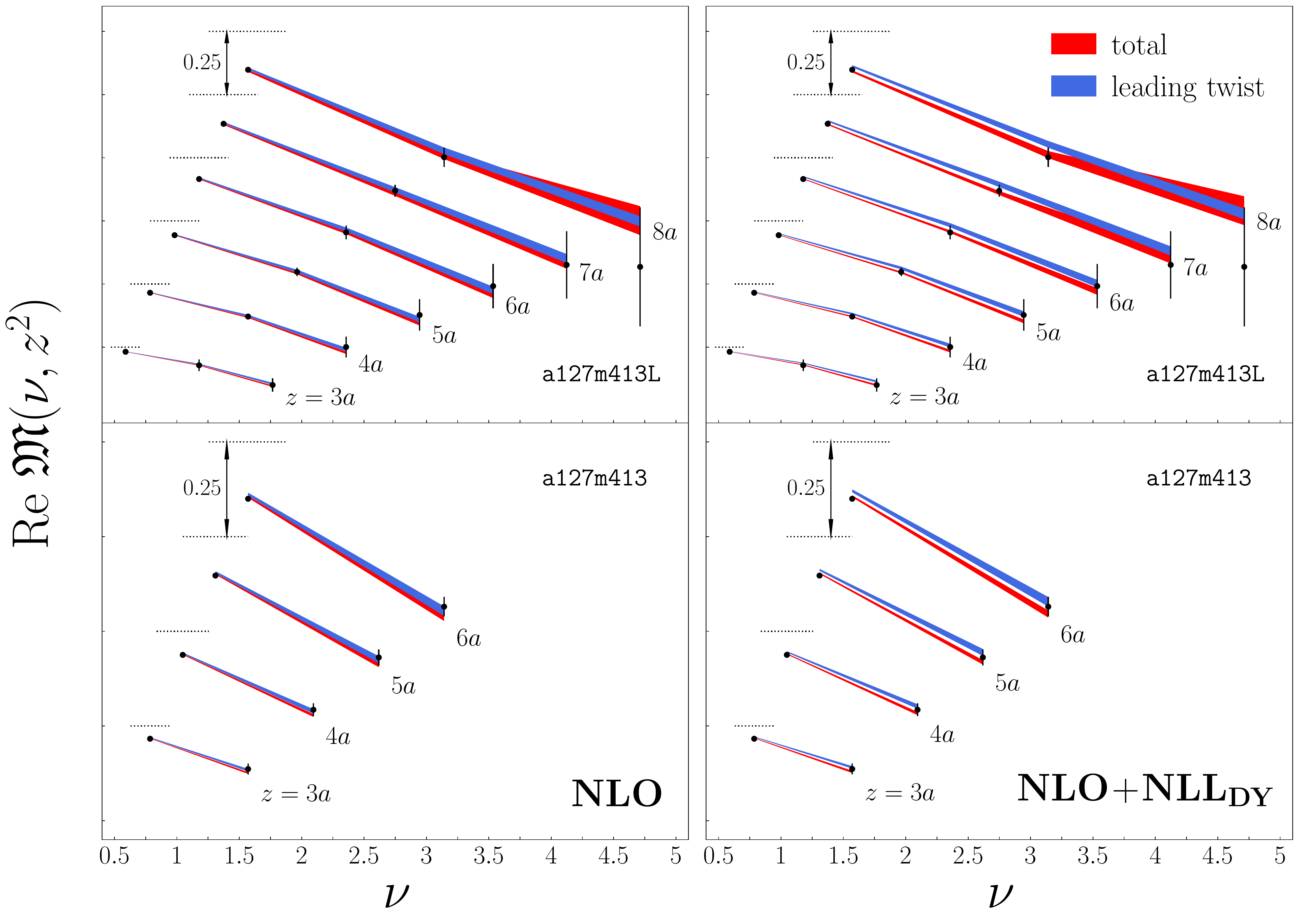}
    \caption{Comparison for the fitted reduced Ioffe time pseudo-distributions using NLO ({\bf left}) and NLO+NLL$_{\rm DY}$ ({\bf right}) methods with the larger volume ({\tt a127m413L}, {\bf top}) and smaller volume ({\tt a127m413}, {\bf bottom}) lattices (black circles).
    The total result of the Bayesian inference 
    (1$\sigma$ red bands) is compared with the leading twist contribution without systematic corrections (1$\sigma$ blue bands). 
    The horizontal dotted lines for each bin in $z$ indicate 1, the normalization value at $\nu=0$.}
    \label{f.RpITD_data}
\end{figure}

In Fig.~\ref{f.RpITD_data} we show the resulting predictions of the Rp-ITD from the analysis, along with the leading twist contributions, with the difference between the bands reflecting the contribution from the systematic corrections.
The panels illustrate the results using the NLO and NLO+NLL$_{\rm DY}$ methods, as well as for the larger volume ($L=32a$, $\tt a127m413L$) and smaller volume ($L=24a$, $\tt a127m413$) lattice ensembles.
Each separated spectrum represents different values of $z$, offset for clarity.
For small values of $\nu$ and $z$, the difference between the total and leading twist bands is effectively zero, which indicates that the systematic corrections do not play a role there, and that the leading power contribution dominates the total lattice observable.
As $\nu$ and $z$ increase, the systematic contributions grow in magnitude, particularly for the NLO+NLL$_{\rm DY}$ case, and are negative, as indicated by the separation between the total and leading twist bands.
In the case of the NLO method, the bands are somewhat separated, but not as pronounced as for the NLO+NLL$_{\rm DY}$ case, implying different systematic corrections in the two analyses.
Because of the difference between the total and leading twist contributions, excluding the systematic corrections would lead to incorrectly extracted PDFs, as the leading twist band with the PDF would have to shift to agree with the data.
The systematic correction terms in Eq.~\eqref{eq.ReRpITDmatching} must therefore be included in the fit.

\begin{figure}
    \centering
    \includegraphics[width=0.9\textwidth]{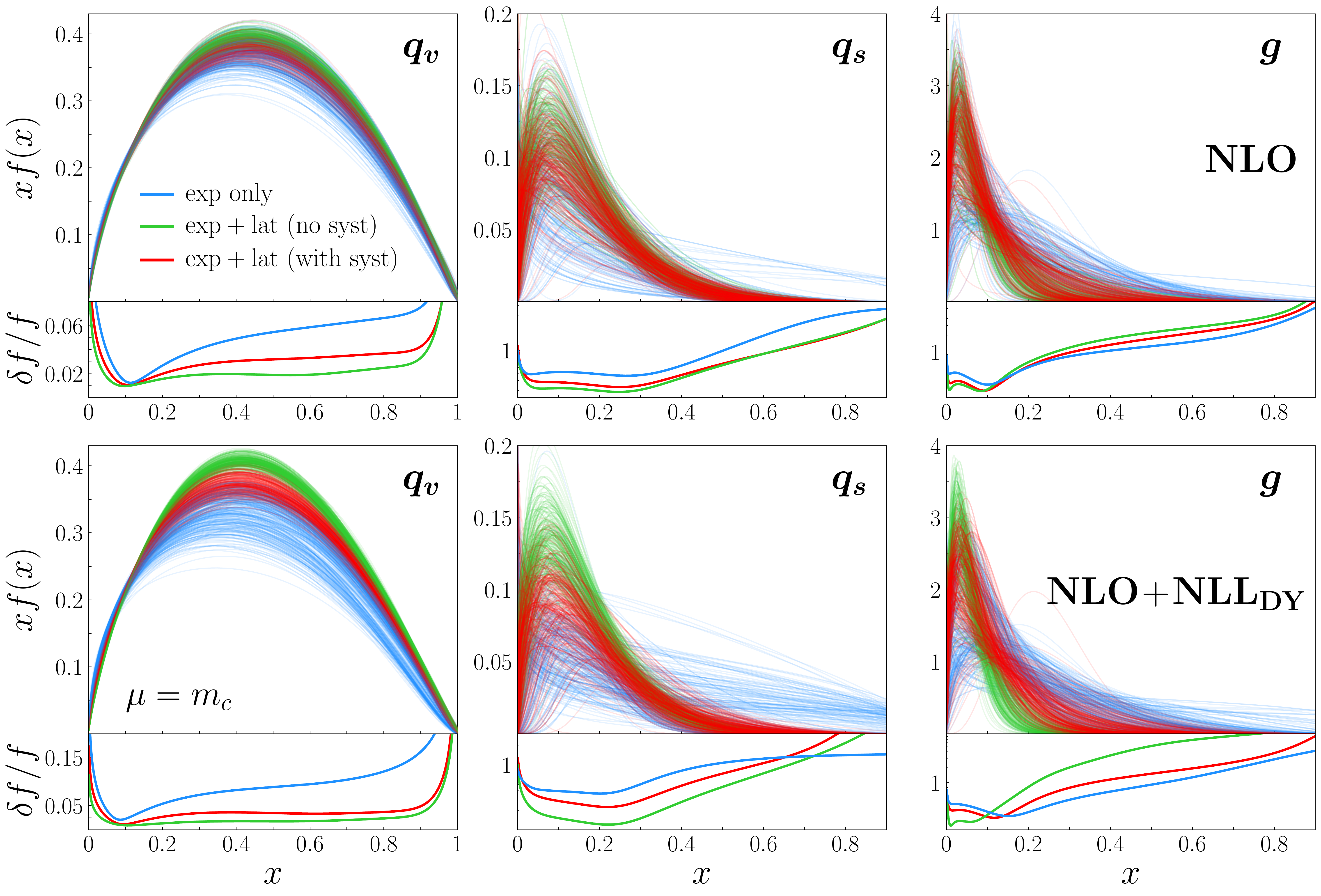}
    \caption{Valence quark ({\bf left}), sea quark ({\bf middle}), and gluon ({\bf right}) distributions with the 1$\sigma$ relative uncertainties (underneath each panel) for the NLO ({\bf top}) and NLO+NLL$_{\rm DY}$ ({\bf bottom}) methods.
    All three scenarios are displayed: 
    Extractions from experimental data alone (Scenario~A, blue curves), from experimental and lattice data without systematic corrections (Scenario~B, green curves), and from including both experimental and lattice data with systematic corrections (Scenario~C, red curves).
    A random subset of 300 of the $\sim 700$ total Monte Carlo replicas is shown.}
    \label{f.RpITD_PDFs}
\end{figure}

The $x$ dependence of the fitted valence quark, sea quark, and gluon distributions, along with the ratios of their uncertainties relative to their mean values, are shown in Fig.~\ref{f.RpITD_PDFs} at the input scale $\mu=m_c$ from Scenarios A, B and C, for both the NLO and NLO+NLL$_{\rm DY}$ cases.
%--NLO PDFs
When using the NLO hard coefficient in DY, the central values of the valence and sea quark distributions are mostly unaffected by the lattice data. 
There is a slight change in the gluon distribution, whereby the inclusion of lattice data decreases its magnitude for $x \gtrsim 0.2$.
However, the general agreement among the distributions indicates that the lattice and experimental data are compatible.

%--NLO+NLL
For the case of the NLO+NLL$_{\rm DY}$ extractions, none of the scenarios are found to match well with each other, suggesting some instability of the PDFs with the inclusion of the lattice data.
The experimental data prefer a valence quark distribution with a slightly smaller magnitude at intermediate $x$. 
When the lattice data are included, the PDF 
increases by $\sim 30\%$ 
in the range $0.2 \lesssim x \lesssim 0.7$.
When including the systematic corrections, on the other hand, the PDF shifts downwards, but still mostly does not overlap with the experimental-only results.
The large-$x$ sea quark and gluon distributions are supressed with the inclusion of the lattice data because of indirect constraints from the momentum sum rule.
Despite the differences of the PDFs among the scenarios, the description of the experimental data remains unchanged, as indicated in Table~\ref{t.chi2_RpITD}.

The PDFs extracted from only the experimental data carry large uncertainties, especially in the NLO+NLL$_{\rm DY}$ case, and including the precise lattice data decreases the uncertainty significantly.
However, including the systematic corrections again increases the uncertainty of the PDFs, because of the increase in the number of parameters, but nevertheless provides a sizable impact.
The behavior of the relative uncertainty in the gluon distribution across the scenarios is opposite to that for the quark distributions, which can be attributed to the redistribution among the parton flavors across the scenarios.

\begin{figure}
    \centering
    \includegraphics[width=0.9\textwidth]{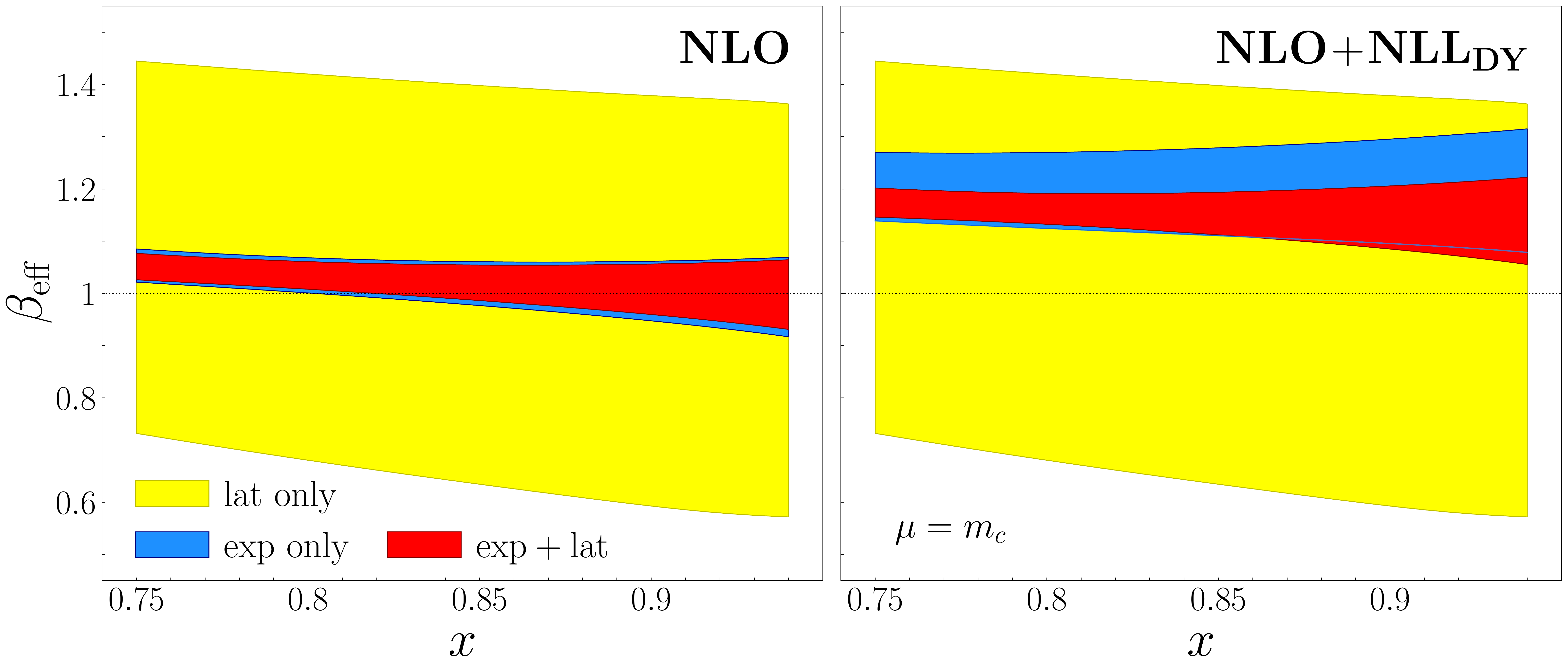}
    \caption{Effective large-$x$ exponent $\beta_{\rm {eff}}$ for the valence quark distribution as a function of $x$ at the input scale $\mu = m_c$ extracted from lattice data alone (yellow bands), experimental data alone (blue bands), and both lattice and experimental data (red bands) from the NLO ({\bf left}) and NLO+NLL$_{\rm DY}$ ({\bf right}) methods.}
    \label{f.betaeff}
\end{figure}

The effective $\beta_v$ parameter describes the degree of falloff at large $x$ in the valence quark distribution, and operationally we define~\cite{Nocera:2014uea, Courtoy:2020fex, Courtoy:2021xpb}
\begin{equation}
\beta_{\rm eff}(x,\mu) = \frac{\partial \log |q_v(x,\mu)|}{\partial \log (1-x)}
    \label{eq.beff}
\end{equation}
at the scale $\mu$.
To obtain the PDF when extracting from lattice data alone, precise data over a large range of $\nu$ is needed.
Jo\'{o} {\it et al.}~\cite{Joo:2019bzr} found $\beta_{\rm eff} \sim 1$, but with a large uncertainty, because of the limited range of $\nu$.
The recent analysis of experimental data in Ref.~\cite{Barry:2021osv} found $\beta_{\rm eff} \sim 1$ with NLO hard coefficients, and $\beta_{\rm eff} \sim 1.2$ when using NLO+NLL$_{\rm DY}$ with double Mellin threshold resummation on the hard coefficients in DY.

In the present analysis, we include the Rp-ITD lattice data and demonstrate in Fig.~\ref{f.betaeff} that the $\beta_{\rm eff}$ resulting from each method of the short distance DY coefficients agrees, within uncertainties, with the $\beta_{\rm eff}$ from using experimental data alone~\cite{Barry:2021osv}.
While $\beta_{\rm eff}$ does not change at all in the NLO case, it is more precisely determined with the inclusion of lattice data for the NLO+NLL$_{\rm DY}$ calculation, as evident from the shrinking of the uncertainty band relative to the experimental only band.
In fact, the $\beta_{\rm eff}$ of the NLO+NLL$_{\rm DY}$ analysis is more centered around the lower portion of the previous band, decreasing slightly its mean value and bringing it closer to $\beta_{\rm eff}=1$.
The difference in uncertainty reduction in the two cases of the hard coefficients can be attributed to the more dramatic decrease in relative uncertainty of the valence quark distribution in the NLO+NLL$_{\rm DY}$ method at large $x$ compared with the NLO case, shown in Fig.~\ref{f.RpITD_PDFs}.

% . . . . . . . . . . . . . . . . . . . . . . . . . . . . . . . . . . . . 
\subsubsection{Which lattice data have maximal impact?}

When studying the impact of adding lattice data to global phenomenological extractions of PDFs, it is pertinent to ask which lattice datasets have the most impact.
In lattice calculations, all hadrons suffer from an exponential growth of the signal-to-noise ratio as energy increases, but the effect on the pion, as the lightest hadron, is most significant. 
This signal-to-noise ratio of lattice correlation functions decays exponentially as $\sim \exp[-(E_h-\frac12 m_\pi n_q)T]$, where $E_h$ is the energy of the hadron, $n_q$ is the number of valence quarks, and $T$ is the Euclidean time separation of the operators \cite{Parisi:1983ae,Lepage:1989hd}.
Excited state contamination is lessened at large time separations, where the pion has a larger signal-to-noise ratio than the nucleon at a fixed low momentum.
In the datasets used here, the relative uncertainty of the lowest momentum state data having $p=1$ (hereafter referred in units of $2\pi/La$) is an order of magnitude smaller that the $p=2$ data, even though all momenta were calculated with equal computational cost.
As such, we expect the lowest momentum state to drive the impact from the lattice data overall. 
This feature also means that utilizing factorization methods that limit power corrections at low momentum is crucial for efficient and precise numerical calculations of parton structure, particularly for the pion, where the effects can be rather dramatic.

As demonstrated with mock data in Ref.~\cite{Karpie:2019eiq}, the range of Ioffe time can dramatically change the quality of the resulting PDF, specifically in the low-$x$ region. 
However, in this combined analysis, the LN data provide constraints at low-$x$ such that an improvement is possible across the whole range of $x$.

For each value of $z$, the smallest $\nu$ values correspond to the $p=1$ points.
As seen in Fig.~\ref{f.RpITD_data}, the leading twist terms generally agree the best with the total results when $\nu$ is small, so that the systematic corrections do not compete with the leading twist contribution.
We~performed a Monte Carlo analysis of the experimental data combined with the $p=1$ lattice points with the same systematic corrections included above, again using both NLO and NLO+NLL$_{\rm DY}$ coefficients.
These analyses included 6 and 4 data points from the $\tt a127m413L$ and $\tt a127m413$ ensembles, respectively.
The resulting $\overline{\chi}_e^2$ are almost identical to Scenario C in Table~\ref{t.chi2_RpITD}, with the NLO analysis producing $\overline{\chi}^2_\lambda = 0.80$ and 1.02 for $\tt a127m413L$ and $\tt a127m413$, respectively, and the NLO+NLL$_{\rm DY}$ giving $\overline{\chi}^2_\lambda = 0.82$ and 1.05 for the same datasets.

\begin{figure}
    \centering
    \includegraphics[width=0.9\textwidth]{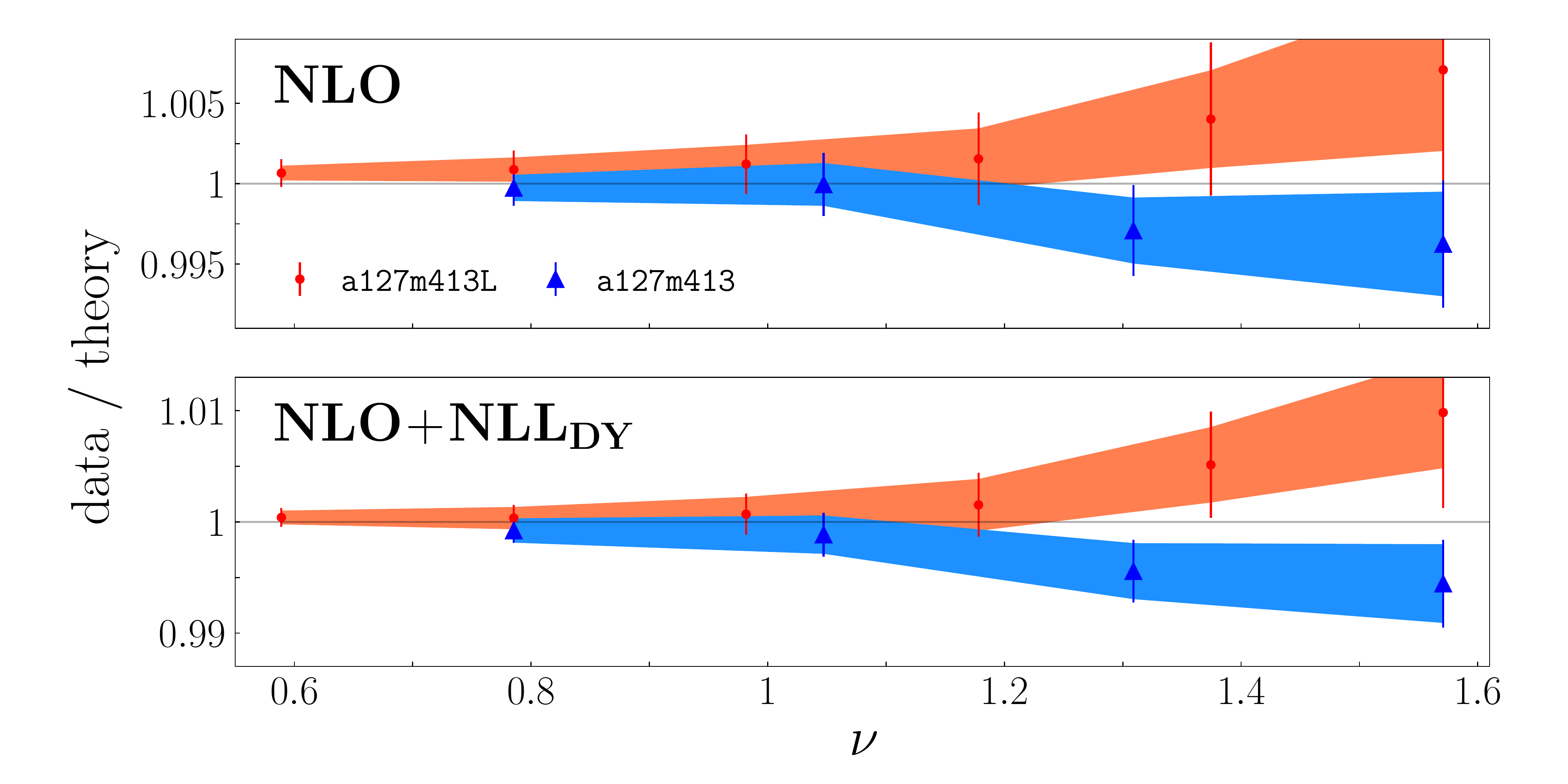}
    \caption{Data to theory ratios for the Rp-ITD datasets $\tt a127m413L$ (red circles) and $\tt a127m413$ (blue triangles) resulting from fits including only the $p=1$ data points for the NLO ({\bf top}) and NLO+NLL$_{\rm DY}$ ({\bf bottom}) methods. The bands represent 1$\sigma$ uncertainties.}
    \label{f.p1datatheory}
\end{figure}

The ratios of the $p=1$ data to the mean values of the theory are shown in Fig.~\ref{f.p1datatheory} for each of the Rp-ITD datasets.
Since all data points have the same momentum $p$, the increase in $\nu$ is directly proportional to the increase in the $z$ spatial separation.
Each of these points is close to unity within the uncertainties of the theory, indicating a good agreement in the analysis.
The datasets agree best with the theory at small $\nu$, and deviate in opposite directions at large $\nu$.
The theory slightly underpredicts the $\tt a127m413L$ data, while marginally overpredicting the $\tt a127m413$ data.
The difference between NLO and NLO+NLL$_{\rm DY}$ is minimal, with a modest preference for the NLO analysis from the $\tt a127m413$ dataset, evident at large $\nu$.

\begin{figure}
    \centering
    \includegraphics[width=0.9\textwidth]{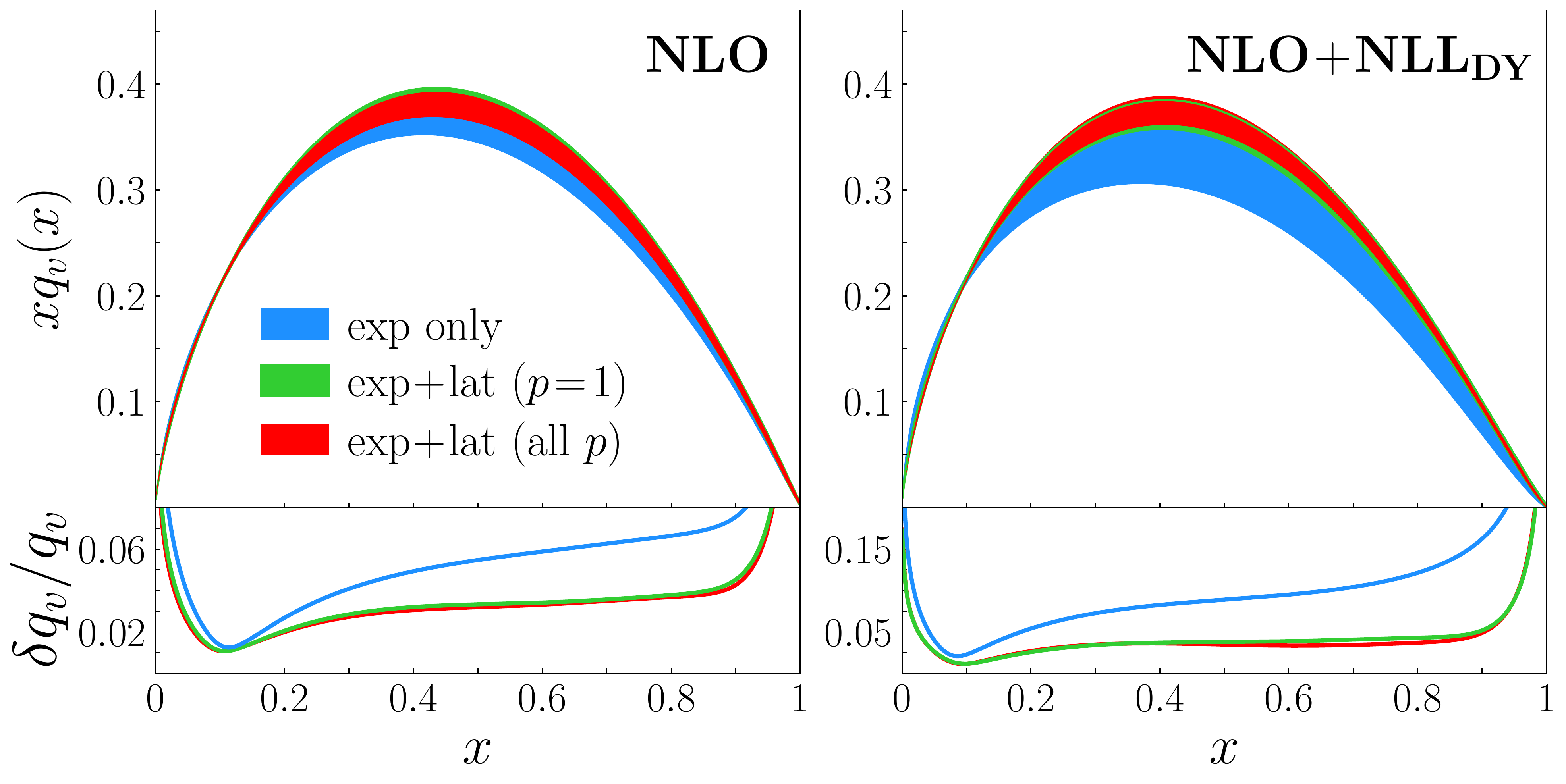}
    \caption{Valence quark distributions ({\bf top}) when extracted from experimental data alone (blue), combined with the $p=1$ lattice data (green), and combined with all the lattice data (red) for the NLO ({\bf left}) and NLO+NLL$_{\rm DY}$ ({\bf right}) cases, along with the relative uncertainties ({\bf bottom}).
    The bands represent a $1\sigma$ uncertainty level.}
    \label{f.pdfp1}
\end{figure}

In comparison with Fig.~\ref{f.RpITD_PDFs}, which showed the PDFs when taking into account all of the momentum values in $\tt a127m413L$ and $p=1,2$ in $\tt a127m413$, in Fig.~\ref{f.pdfp1} we illustrate the resulting valence quark distribution when including only the $p=1$ lattice data points.
The PDFs inferred from the combined analysis that includes all the lattice data and their systematic corrections are almost indistinguishable from those inferred from the same combined analysis but only including the $p=1$ lattice data points.
As expected, the relative uncertainties decrease when including more lattice data points, but only marginally.
The results indicate that the $p=1$ points drive the analysis from the lattice data and provide the strongest constraints on the PDFs.
Despite the limited Ioffe time range of the lattice data, there is clearly merit in performing extractions from the lowest momentum values, as these are the most precise and produce the same qualitative and quantitative results as the analysis with all momenta.

% . . . . . . . . . . . . . . . . . . . . . . . . . . . . . . . . . . . . 
\subsubsection{Scale variation}

\begin{figure}
    \centering
    \includegraphics[width=0.9\textwidth]{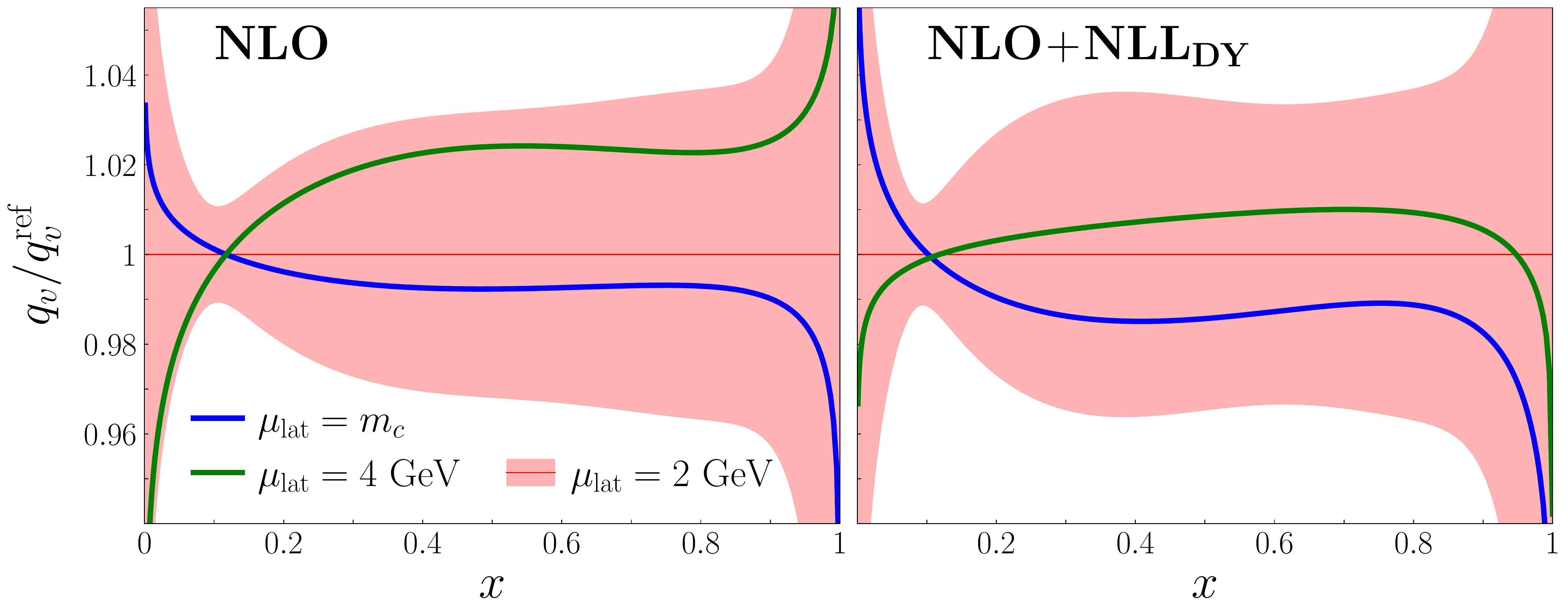}
    \caption{Central values of the extracted pion valence quark distribution at the input scale from fits setting $\muL=m_c$ (blue lines), $\muL=4~{\rm GeV}$ (green lines) relative to $q_v^{\rm ref}$, which is the central value of the $\muL=2~{\rm GeV}$ main result (red lines and $1\sigma$ pink uncertainty bands) for the NLO ({\bf left}) and NLO+NLL$_{\rm DY}$ ({\bf right}) cases.}
    \label{f.pdfscale}
\end{figure}

Following Sec.~\ref{ssec.scale}, we investigate here in detail the effects of varying the scale, $\muL$.
In addition to the main analysis, which uses $\muL=2~{\rm GeV}$, we performed two further combined analyses, varying the scales to $\muL=m_c$ and $\muL=4~{\rm GeV}$, using all momenta from the lattice data and the same systematic terms as in Eq.~(\ref{eq.ReRpITDmatching}).
The central values of the valence quark distributions extracted using the input scales $\muL=m_c$ and $\muL=4~{\rm GeV}$ are shown in Fig.~\ref{f.pdfscale} relative to the mean value of the $\muL=2~{\rm GeV}$ result, for both of the calculational methods for the short-distance DY coefficients.

We find that the $\muL=m_c$ and $\muL=4~{\rm GeV}$ versions of the analysis slightly deviate from the $\muL=2~{\rm GeV}$ results, but only by $\lesssim 2\%$ for most of the accessible $x$ range in each case.
Importantly, each of the central values lie within the 1$\sigma$ uncertainty band from the $\muL=2~{\rm GeV}$ result.
This suggests that the uncertainty band associated with the valence quark PDF has not been underestimated due to scale variation effects, and demonstrates that these effects for the Rp-ITD are largely insignificant.

% . . . . . . . . . . . . . . . . . . . . . . . . . . . . . . . . . . . .
\subsubsection{Quantification of Rp-ITD systematics}
\label{ssec.syst}

\begin{figure}
    \centering
    \includegraphics[width=0.8\textwidth]{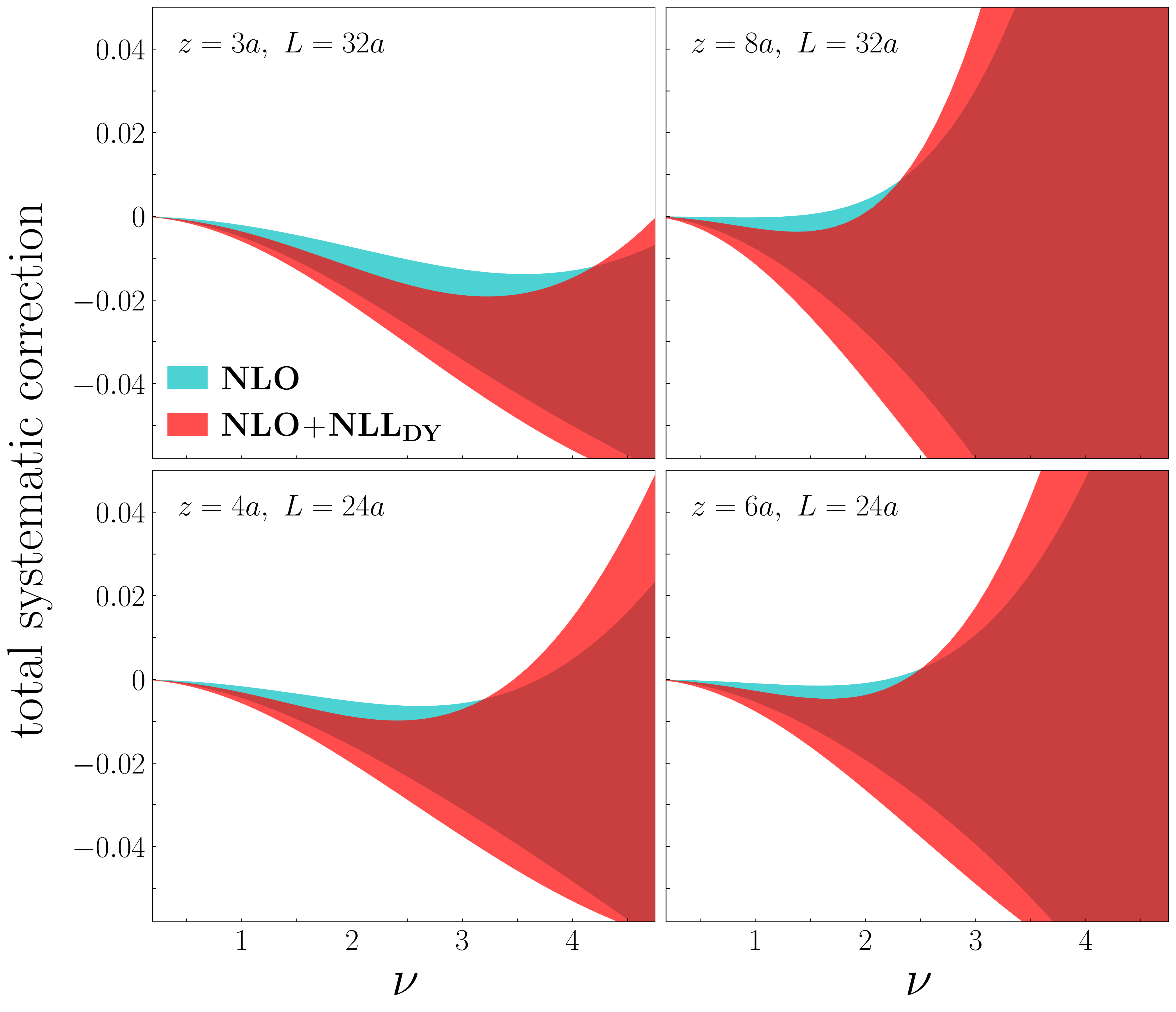}
    \caption{Total systematic correction versus Ioffe time $\nu$ from the NLO (cyan bands) and NLO+NLL$_{\rm DY}$ (red bands) extractions at several representative values of $z$ and $L$ in the available Rp-ITD datasets.  The bands represent $1\sigma$ uncertainty levels.}
    \label{f.syst_RpITD}
\end{figure}

An important aspect of our analysis is the ability to use the lattice QCD data to provide information on both the pion PDFs and the systematic uncertainties on the lattice calculations.
The leading twist contributions include dependence on the PDFs already constrained by the experimental data, whereas the systematic corrections are {\it a priori} unknown.
In Fig.~\ref{f.RpITD_data}, above, we showed that for small $\nu$, $p$, and $z$ the leading twist contribution largely equaled the total theory prediction, indicating that the systematic corrections in Eq.~(\ref{eq.ReRpITDmatching}) are small.
The NLO method tended to give smaller deviation between the two curves, while the NLO+NLL$_{\rm DY}$ implied a somewhat more negative systematic correction.
In Fig.~\ref{f.syst_RpITD}, we show the total systematic corrections associated with the lattice from extractions of both the NLO and NLO+NLL$_{\rm DY}$ analyses.
When using the NLO theory for the hard coefficients in the DY calculation, the systematic corrections agree well with those using the NLO+NLL$_{\rm DY}$ theory in both magnitude and uncertainty.

There is considerable overlap between the two sets of bands, and the total systematic corrections are almost indistinguishable.
New data are needed to futher discriminate between the methods and provide a more trustworthy description of the systematic corrections on these lattices.
The general trend between the two methods is common: the systematic corrections are small when $\nu \approx 0$, and increase in magnitude and uncertainty with $\nu$.
The full systematic corrections as functions of $\nu$ tend to be negative with a downward slope up to a minimum, after which the systematic corrections increase.
As $z$ increases, the minimum appears at smaller $\nu$.
Additionally, both methods indicate the uncertainty of the systematic corrections increases with $z$.
Avoiding the lattice systematic corrections and performing calculations that focus on regions in which the leading twist contribution dominates can effectively be done at small~$\nu$.
The systematic uncertainty bands shown in Fig.~\ref{f.syst_RpITD} are either comparable or larger than the statistical uncertainties shown by the lattice data in Fig.~\ref{f.RpITD_data}.
To further constrain the PDFs from these data, an improvement should be performed not only on the statistics, but also on the systematics, especially at larger values of $\nu$.

\begin{figure}
    \centering
    \includegraphics[width=0.95\textwidth]{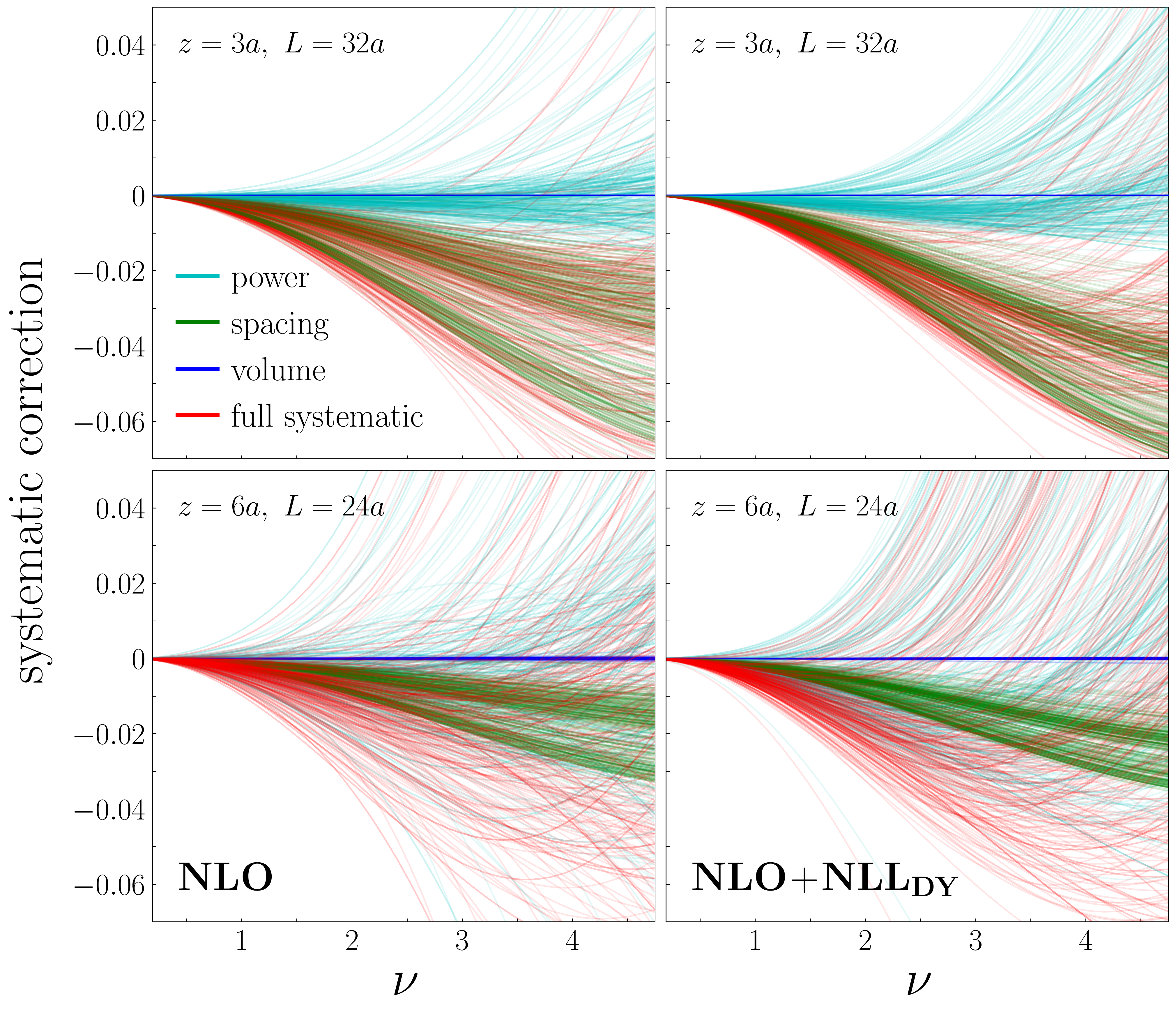}
    \caption{Contributions of the power (cyan), lattice spacing (green), and finite volume (blue) systematic corrections as in Eq.~(\ref{eq.ReRpITDmatching}), along with the sum (red), for the ({\bf top}) $\tt a127m413L$ and ({\bf bottom}) $\tt a127m413$ ensembles at various values of the lattice separation $z$ for the NLO ({\bf left}) and NLO+NLL$_{\rm DY}$ ({\bf right}) methods.
    Each panel is evaluated at $m_\pi=413$~MeV and $a=0.127$~fm, and for clarity a subset of 300 replicas is shown from a total of $\approx 720$.}
    \label{f.syst_reps_RpITD}
\end{figure}

In Fig.~\ref{f.syst_reps_RpITD} we show replicas from the Monte Carlo analysis of the systematic corrections for the power, lattice spacing, and finite volume correction terms in Eq.~\eqref{eq.ReRpITDmatching}.
These corrections are shown for $z=3a$ and $z=6a$, for both lattice volumes, and for both calculations of the short distance coefficients in DY.
At small $z$ values, the power correction terms are quite small and the total systematic terms are dominated by the lattice spacing errors.
However, at large~$z$, the power corrections terms play a much more important role, while the lattice spacing errors have a lesser impact on the total systematics.
The finite volume corrections in both cases are effectively zero and provide no contribution to the overall systematic corrections.
Even though we may test the sensitivity to finite volume corrections because of the different lattice volumes, the contribution is negligible compared with the power and spacing corrections, albeit slightly larger for the smaller lattice volume.

%.....................................................................
\subsection{Current-current correlator analysis}

In this section we present results obtained by combining CC correlators as in Eq.~(\ref{eq.CCcorrelator}) with experimental data in our global analysis.
The statistical uncertainties on the lattice data for CC correlators~\cite{Sufian:2020vzb} are somewhat larger than for the Rp-ITD data, which suggests that the CC correlator data may not constrain the PDFs  as strongly as do the Rp-ITD data.
However, it is nevertheless important to quantify the extent to which the existing CC correlator data impact the extraction of the pion PDFs, and gauge what kind of uncertainties future lattice simulations should aim for.

For the CC correlator observable, we consider the $\chi^2_e$ as in Eq.~\eqref{eq.chi2expt} for each $\chi^2$ minimization and use $\overline{\chi}^2_e$ as in Eq.~\eqref{eq.rchi2e} for analysis purposes.
That is, we do not consider the covariance matrix and the correlation between individual uncertainties.
We use the systematic correction terms associated with the power corrections and lattice spacing errors, as shown in Eq.~\eqref{eq.CCcorrelator}.
We found that by excluding all systematic effects, there was tension between the experimental and lattice data, evident in the resulting PDFs, which accepted a small set of solutions that were noticeably different from the experimental-only analysis.
Additionally, including more systematic corrections increased the uncertainties on the Drell-Yan predictions generated from the experimental-only analysis.
These inconsistencies led us to consider various systematic effects, and incorporating power and discretization corrections tamed both the uncertainties on the Drell-Yan predictions and tensions in the PDFs.

In Table~\ref{t.chi2_cc} we show the reduced $\chi^2$ for each of the datasets in the analysis.
The agreement of the resulting theory with the DY and LN data does not change significantly from the experimental-only results shown as Scenario A in Table~\ref{t.chi2_RpITD}.
The overall $\overline{\chi}^2=0.81$ and $0.80$ for the NLO and NLO+NLL$_{\rm DY}$ methods, respectively, indicate a good agreement with the data as a whole, as well as with individual experimental and CC datasets, the only exception being the {\tt a127m413} ensemble that has $\overline{\chi}^2 \sim 1.9$ for each of the methods.
Noticeably, the $\overline{\chi}^2$ values for each method are very close, indicating that the CC data do not prefer one method over the other.

\begin{table}
\centering
\caption{Summary of results for our global analysis of the DY and LN electroproduction data, along with the CC correlator lattice data, including the number of data points fitted $N_{\rm dat}$ and the $\overline{\chi}^2$ values for each of the calculations of the DY hard coefficient. The CC correlator data were fitted with $\muL=2~{\rm GeV}$.\\}
\begin{tabular}{llc|c|c}
\hline
&&& ~~~NLO~~~&~~~NLO+$\rm NLL_{DY}$~~~\\

~Process~~ & ~Experiment~~  
        & ~~~$N_{\rm dat}$~~~  
        & ~~~$\overline{\chi}^2$~~~
        & ~~~$\overline{\chi}^2$~~~\\
\hline
{\bf ~DY}& ~E615 $(x_F,Q)$          & 61
                                    & 0.84
                                    & 0.81\\
         & ~NA10~\mbox{\scriptsize (194~GeV)} $(x_F,Q)$
                                    & 36
                                    & 0.53
                                    & 0.54\\
         & ~NA10~\mbox{\scriptsize (286~GeV)} $(x_F,Q)$ 
                                    & 20
                                    & 0.81
                                    & 0.85\\
{\bf ~LN}& ~H1                      & 58
                                    & 0.37
                                    & 0.35\\
         & ~ZEUS                    & 50
                                    & 1.55
                                    & 1.54\\ 
{\bf ~CC}     & ~\tt a94m278        & 20
                                    & 0.33
                                    & 0.33\\
              & ~\tt a94m358        & 20
                                    & 0.47
                                    & 0.46\\
              & ~\tt a127m413L      & 12
                                    & 0.72
                                    & 0.74\\
              & ~\tt a127m413       & 12
                                    & 1.92
                                    & 1.91\\ \hline
~\textbf{Total} &                   & {\bf 289}
                                    & {\bf 0.81}
                                    & {\bf 0.80}  \\ \hline \\
\end{tabular}

\label{t.chi2_cc}
\end{table}

\begin{figure}[t]
    \centering
    \includegraphics[width=0.975\textwidth]{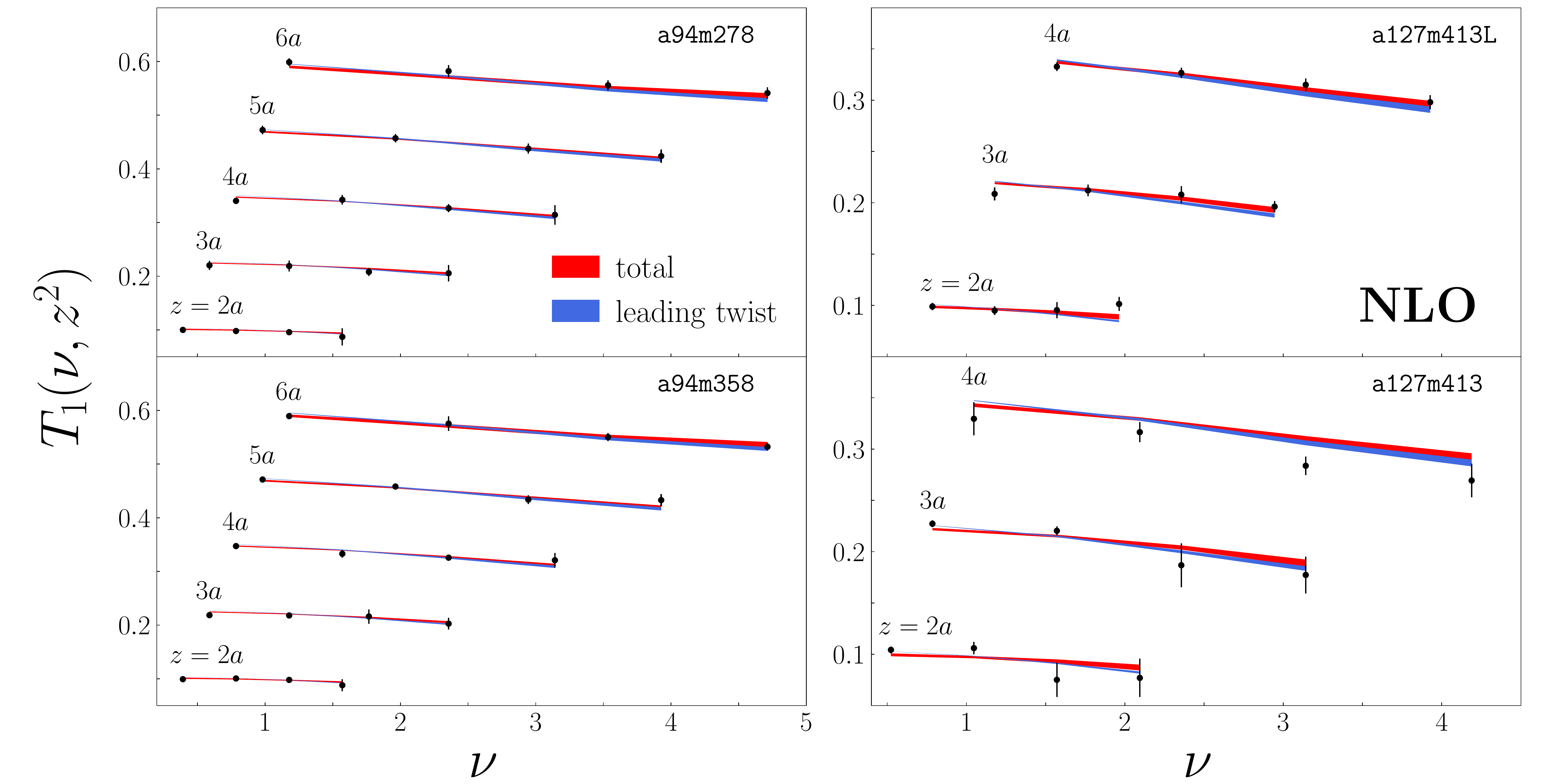}
    \vspace*{-0.5cm}
    \caption{CC correlator data versus Ioffee time $\nu$ for the datasets for the full theory as in Eq.~\eqref{eq.CCcorrelator} (red bands with $1\sigma$ uncertainties) and the leading twist contributions (blue bands) for the NLO method.  An additive shift of $(z/a-2)/8$ beyond $z=2a$ is applied for clarity.\\}
    \label{f.CC_NLO_data}
\end{figure}

\begin{figure}[h]
    \centering
    \vspace*{-0.5cm}
    \includegraphics[width=0.975\textwidth]{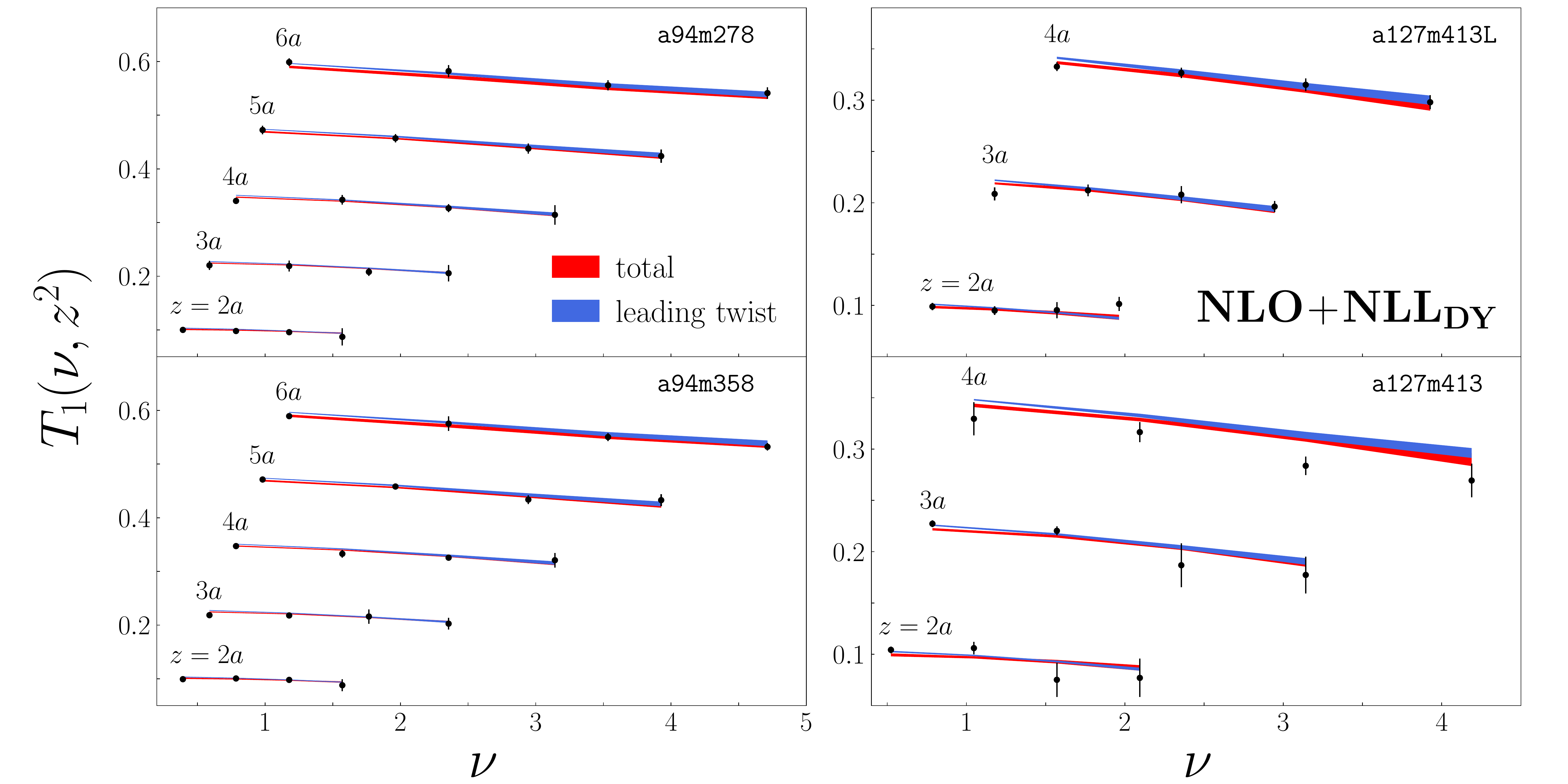}
    \caption{Same as Fig.~\ref{f.CC_NLO_data}, but for the NLO+NLL$_{\rm DY}$ method.}
    \label{f.CC_NLONLL_data}
\end{figure}

In Figs.~\ref{f.CC_NLO_data} and \ref{f.CC_NLONLL_data} we compare the full theory and the leading twist contribution with the lattice data for the NLO and NLO+NLL$_{\rm DY}$ methods, respectively.
For this observable, the uncertainties are large in comparison with the Rp-ITD data, especially at small momentum.
While the leading twist and total contributions have considerable overlap, notably in the lattice data with $a=0.094$~fm, various features of the data are difficult to capture by using next-to-leading order leading power matching coefficients, and systematic corrections are needed to achieve good agreement with the full set of lattice data.

\begin{figure}[h]
    \centering
    \includegraphics[width=0.95\textwidth]{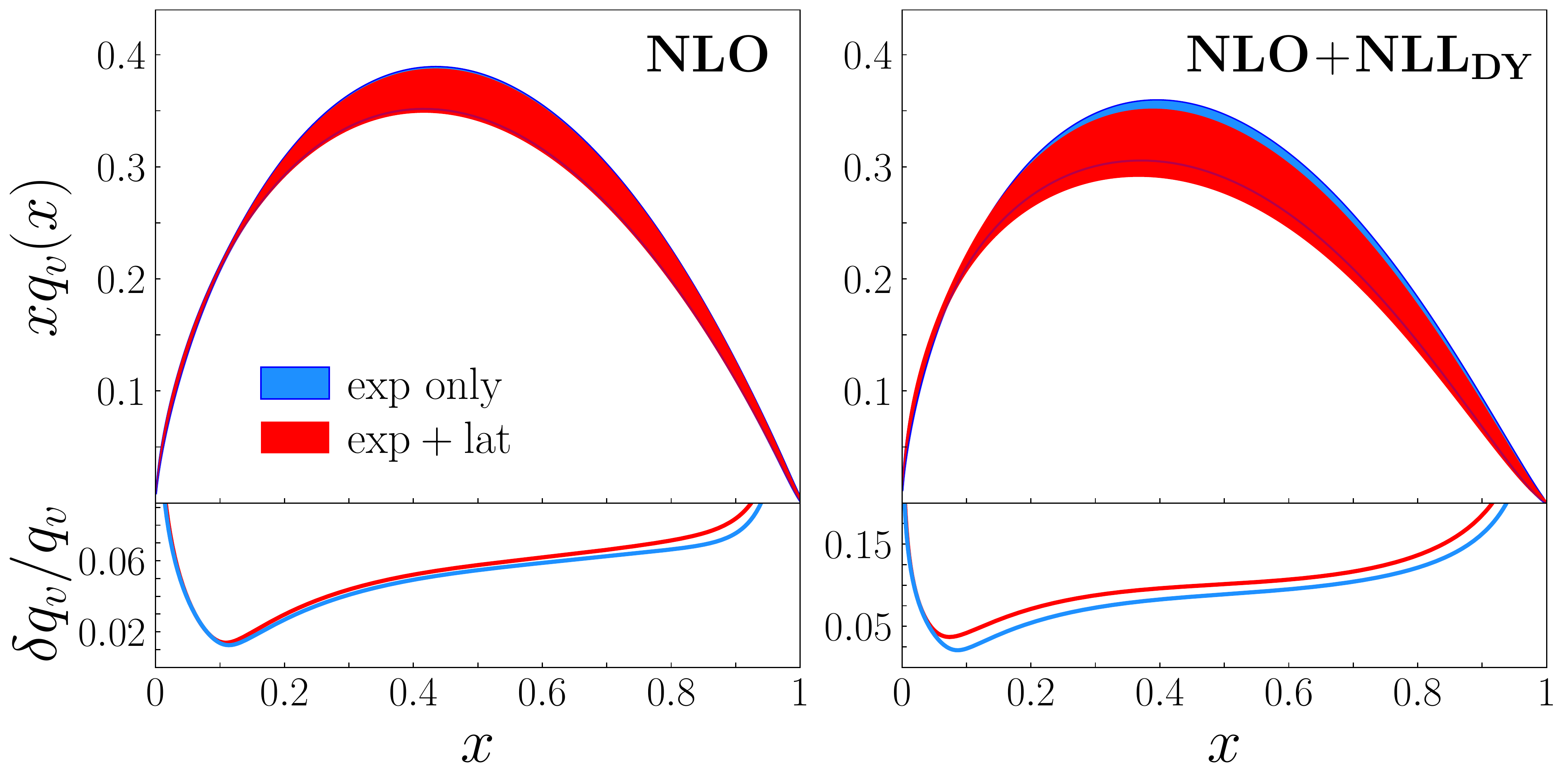}
    \caption{Valence quark distribution $x q_v(x)$ ({\bf top}) and the relative uncertainty $\delta q_v/q_v$ ({\bf bottom}) from the experimental-only analysis (blue $1\sigma$ uncertainty bands) and with inclusion of the CC lattice data (red bands) for the NLO ({\bf left}) and NLO+NLL$_{\rm DY}$ ({\bf right}) methods.}
    \label{f.PDFs_CC}
\end{figure}

The valence quark distributions before and after the inclusion of the lattice CC data are shown in Fig.~\ref{f.PDFs_CC} for the NLO and NLO+NLL$_{\rm DY}$ methods, along with their relative uncertainties.
The PDFs themselves are rather stable under the inclusion of the lattice data, suggesting compatibility between these lattice and experimental datasets.
Unlike the Rp-ITD data, the CC lattice data have essentially no pull on the central values for the NLO+NLL$_{\rm DY}$ analysis, indicating that these data have minor impact on the PDFs.
The relative uncertainties are very similar in both cases, with a slight increase when including the CC lattice data compared with the experimental-only analysis.
Similar behaviors are seen for the sea quark and gluon distributions (not shown here).
Despite the small increase in the relative uncertainties of the PDFs, the relative uncertainties on the predictions for the experimental observables using these PDFs do not increase, which is a reflection of the nontrivial interplay between different flavors of PDFs.

\begin{figure}[h]
    \centering
    \includegraphics[width=0.9\textwidth]{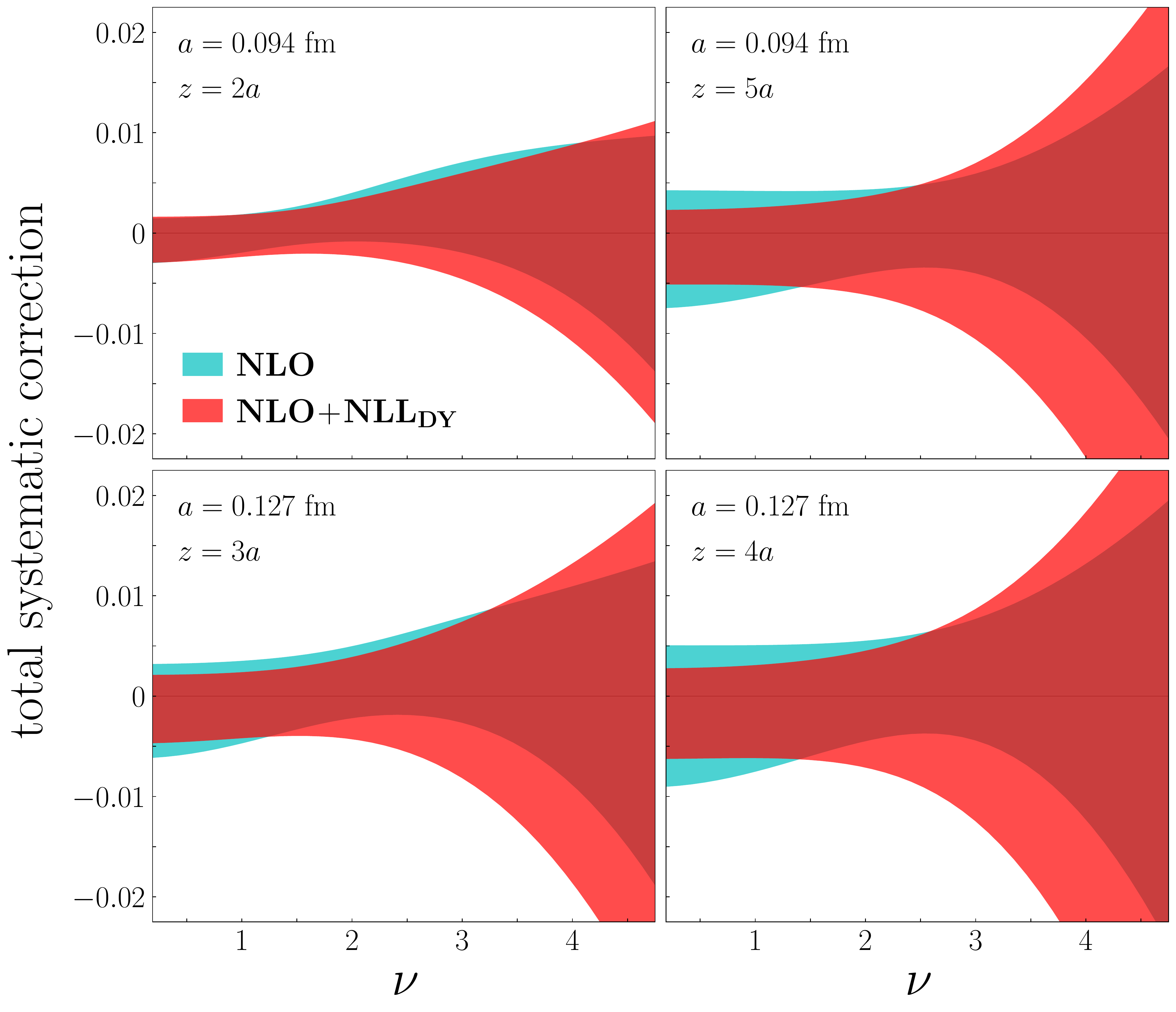}
    \caption{Total systematic corrections from the CC correlators for $a=0.094$~fm ({\bf top}) and $a=0.127$~fm ({\bf bottom}) from the NLO (cyan 1$\sigma$ bands) and NLO+NLL$_{\rm DY}$ (red bands) analyses.}
    \label{f.syst_CC}
\end{figure}

Figure~\ref{f.syst_CC} shows the total systematic corrections from the global analyses for various $z$ and $a$ values that are sensitive to the discretization and power corrections.
Again, considerable overlap occurs between the NLO and NLO+NLL$_{\rm DY}$ results, as was similarly observed in the Rp-ITD analysis.
Contrary to the Rp-ITD analysis, the systematic corrections at the smallest $\nu$ are not necessarily zero by construction, and in fact are slightly negative.
In the NLO case, the total correction generally increases as $\nu$ increases for all values of $z$ and $a$, and at large $\nu$ trends downwards.
While the corrections from using the NLO+NLL$_{\rm DY}$ method follows similar trends as in the NLO analysis, the systematic corrections tend to have larger uncertainties at larger $\nu$.

Even though the existing CC correlator data do not impact the PDFs significantly, the consistency of the extracted PDFs from such a different hadronic matrix element clearly signals the non-trivial success of QCD factorization and universality of PDFs.
We also explore what may be needed in future in order to achieve an impact.
In the current analysis, the total uncertainty associated with the systematic corrections is generally similar to, or larger than, the uncertainty of the total theory shown in Figs.~\ref{f.CC_NLO_data} and \ref{f.CC_NLONLL_data}.
It is unlikely that reducing statistical uncertainties will significantly improve the uncertainty on the PDF, because of the size of the systematic corrections.
We have carried out an impact study in which the statistical uncertainties of the CC lattice data are decreased slowly to match the systematic uncertainty.
As expected, the impact on the PDFs was minimal.
Reducing the uncertainties on the PDFs from the CC correlator data clearly requires in the first instance a better understanding of the systematic corrections.

\begin{figure}
    \centering
    \includegraphics[width=0.9\textwidth]{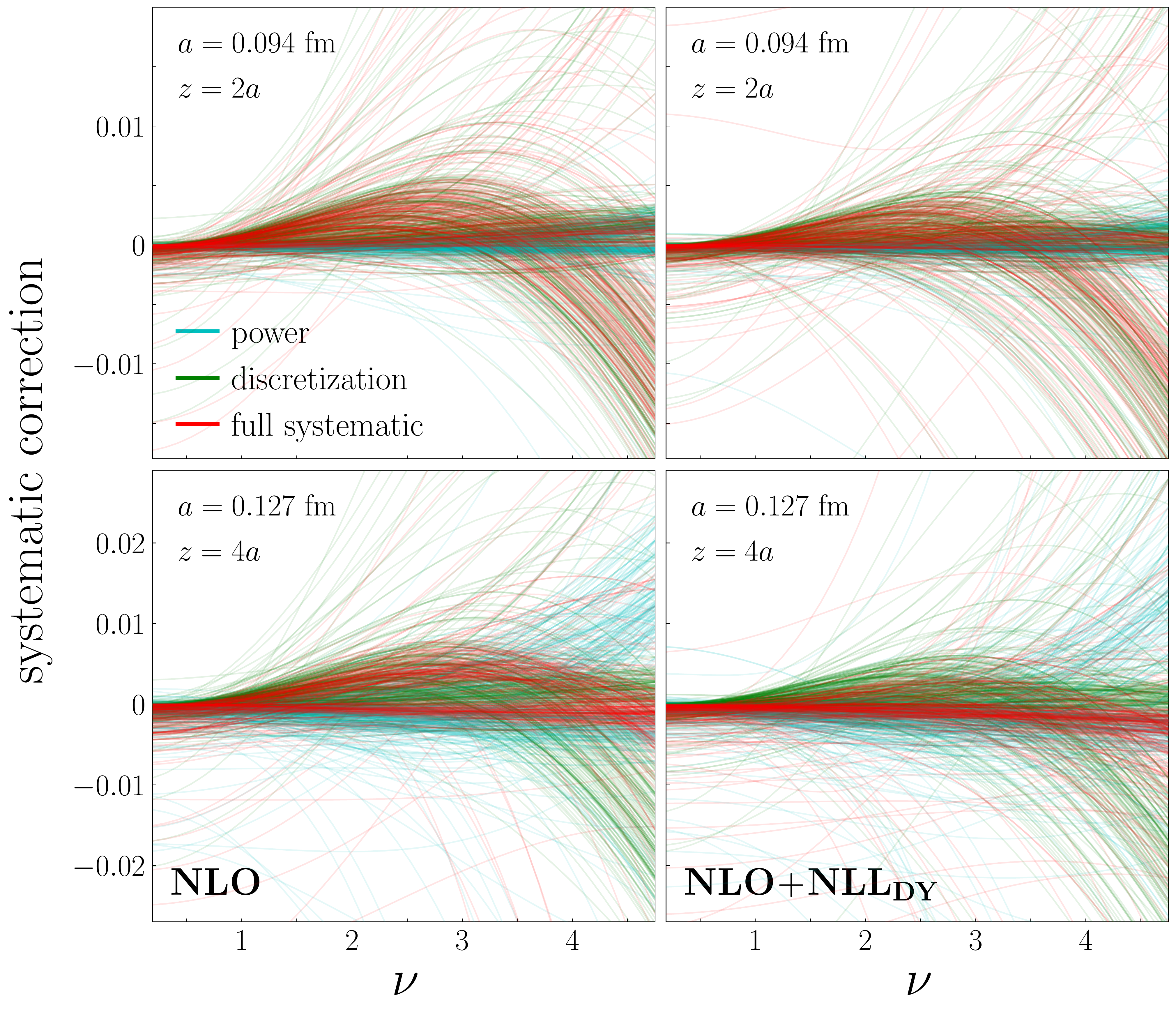}
    \caption{Systematic corrections associated with the CC correlator lattice data broken down into power (cyan), discretization (green), and total (red) corrections from the NLO ({\bf left}) and NLO+NLL$_{\rm DY}$ ({\bf right}) analyses at various $a$ and $z$ values. Shown is a representative subset of 500 replicas.}
    \label{f.syst_reps_CC}
\end{figure}

Finally, in Fig.~\ref{f.syst_reps_CC} we present the breakdown of the systematic corrections in terms of the discretization and power corrections for the NLO and NLO+NLL$_{\rm DY}$ analyses.
As expected, the power corrections are quite small, just away from zero when $z=2a$, but increase in magnitude for $z=4a$.
The discretization corrections are responsible for the general trends of the full systematic uncertainty at small $z$, but these corrections seem to compete with the power corrections at large $z$ and $a$.
When both $z$ and $a$ are large, there tends to be a tension between the two systematic corrections, resulting in a large uncertainty of the total correction, and the full systematic correction is consistent with zero with a large uncertainty.
Further lattice calculations, at different pion masses and lattice spacings and with improved statistical precision, are needed to tame these systematic effects and to test our {\it ansatz} for the correction terms.

%%%%%%%%%%%%%%%%%%%%%%%%%%%%%%%%%%%%%%%%%%%%%%%%%%%%%%%%%%%%%%%%%%%%%%
\section{Conclusion} \label{s.conclusions}

This is the first analysis of its kind, in which experimental data on high-energy pion-nucleus Drell-Yan and leading neutron electroproduction reactions have been supplemented by lattice QCD data on reduced Ioffe time pseudo-distributions and matrix elements of current-current correlators to constrain the PDFs of the pion. 
One of the main motivations of simultaneously fitting both the experimental and lattice data has been to rigorously quantify the uncertainties on the pion PDFs and identify systematic effects intrinsic to the lattice QCD observables.

The use of the NLO or NLO+NLL short distance coefficients for the DY data is not distinguishable on the basis of goodness of fit to the experimental data.
Including the lattice QCD data had {\it a priori} unknown systematic corrections associated with the lattice ensembles, complicating the ability to distinguish between the two DY hard coefficient calculations.
The agreement with the lattice data was similar for the two methods; however, the resulting PDFs were affected in the case of the NLO+NLL$_{\rm DY}$ analysis when including Rp-ITD data.
This suggests the need to further investigate possible tensions between the experimental and lattice observables, and the exercise would benefit from new observables being available.
Regardless of this, the lattice data, in conjunction with the experimental measurements, consistently prefer an effective exponent $\beta_{\rm eff} \approx 1.0-1.2$.

The improvement found in the PDF uncertainties with the inclusion of the Rp-ITD lattice data has implications for future global analyses with combined experimental and lattice QCD observables.
First, since the $p=1$ data have such a limited range of Ioffe time, the Fourier transform and the nature of the inverse problem complicate the determination of an $x$-space quasi-PDF or pseudo-PDF from the lattice data alone, particularly in the low-$x$ region. 
The precise low-$\nu$ data from the lattice can provide a significant constraint on the lowest moments of the PDFs, complemented by the experimental range of data in the low-$x$ region.
The $p=1$ data with ${\cal O}(0.1\%)$ statistical precision seem necessary for providing a useful constraint on the PDFs, while adding the $p=2$ data with ${\cal O}(1\%)$ statistical precision did not dramatically improve the result.
This fact suggests that future lattice calculations which aim to complement experimental datasets will require sub-percent level precision.
These types of constraints can be provided by the lattice data through a factorization method which promotes the use of the low momentum data, such as the pseudo-PDF approach.

Through the complementarity of the lattice and experimental data, we were able to quantify the systematic corrections associated with these ensembles and their uncertainties.
Decreasing the statistical uncertainties on lattice observables is not sufficient to improve our knowledge of pion PDFs, since statistical and systematic effects are comparable in size.
Improved control over systematic effects and reduced statistical uncertainties are both required to further constrain the PDFs.
Future lattice calculations performed at smaller lattice spacings will further limit the impact of the power and discretization corrections.
Here, the leading twist contributions are dominant, allowing the lattice data to isolate the PDFs more cleanly and provide reliable tests of universality on the PDFs.
Finite volume effects were shown to be insignificant for the available Rp-ITD data, so that sampling different lattice volumes may not be essential.
On the experimental side, more observables are needed in regions of kinematics that overlap with existing data in order to test the universality of the pion PDFs.

The importance of combining experimental and lattice QCD data was also evident in analyzing the Rp-ITD data with the smallest momentum.
The lattice data had a significant impact, despite the limited range of Ioffe time, because of the complementarity of the lattice and experimental data.
While this was evident in this analysis with collinear pion PDFs, it should also be noted that this improvement may not be as significant when one considers the nucleon.
Because the mass of the nucleon is larger than that of the pion, the statistical noise of low momentum nucleon lattice QCD correlation functions is generally higher than the pion, though the signal-to-noise ratio of calculations, at fixed computational cost, does not seem to decay as dramatically with momentum as for the pion. 
Additionally, the collinear isovector nucleon PDFs are already quite well constrained by experimental data, so significantly more precise lattice data may be needed to achieve the relative improvement.
Extensions towards nucleon PDFs that are not well constrained by experimental data such as helicity and transversity PDFs could be useful following the methodology presented in this work. \\

%%%%%%%%%%%%%%%%%%%%%%%%%%%%%%%%%%%%%%%%%%%%%%%%%%%%%%%%%%%%%%%%%%%%%%
\paragraph*{Acknowledgements.}

This work is supported by the US Department of Energy (DOE) Contract No.~DE-AC05-06OR23177, under which Jefferson Science Associates, LLC operates Jefferson Lab, and within the framework of the TMD Collaboration.  
We acknowledge the facilities of the USQCD Collaboration used for this research in part, which are funded by the Office of Science (OS) of the US DOE.
This material is based in part upon work supported by a grant from the Southeastern Universities Research Association (SURA) under an appropriation from the Commonwealth of Virginia.
This work used the Extreme Science and Engineering Discovery Environment (XSEDE), which is supported by National Science Foundation (NSF) grant number ACI-1548562~\cite{xsede}.
This work was performed in part using computing facilities at William and Mary which were provided by contributions from the NSF (MRI grant PHY-1626177), the Commonwealth of Virginia Equipment Trust Fund and the Office of Naval Research.
This work used resources at NERSC, a DOE OS User Facility supported by the OS of  the US DOE under Contract~\#DE-AC02-05CH11231, as well as resources of the Oak Ridge Leadership Computing Facility at ORNL, which is supported by the OS of the US DOE under Contract No.~\#DE-AC05-00OR22725. 
The authors gratefully acknowledge the computing time granted by the John von Neumann Institute for Computing (NIC) and provided on the supercomputer JURECA at J\"{u}lich Supercomputing Centre (JSC)~\cite{jureca}. 
K.O. and R.S.S. acknowledge support in part  by the US DOE through Grant Number DE-FG02-04ER41302, by STFC consolidated grant ST/P000681/1. 
The work of N.S. was supported by the DOE, OS, Office of Nuclear Physics in the Early Career Program.
The work of J.K. was supported in part by US DOE grant \#DE32 SC0011941.

\bibliography{literature}

\end{document}